\RequirePackage{fixltx2e}
\documentclass[twocolumn,floatfix,showpacs,preprintnumbers,nofootinbib,%
superscriptaddress]{revtex4}
\usepackage[english]{babel}
\usepackage[utf8x]{inputenc}
\usepackage{mathrsfs}
\usepackage{amssymb}
\usepackage{amsmath}
\usepackage{mathtools}
\usepackage{mathrsfs}
\usepackage{graphicx}
\usepackage[caption=false]{subfig}
\usepackage{fixltx2e}
\usepackage{mciteplus}

\usepackage{xcolor}
\definecolor{lcolor}{rgb}{0.5,0,0}
\definecolor{citcolor}{rgb}{0,0.3,0.0}

\usepackage[breaklinks,colorlinks,urlcolor=blue,citecolor=citcolor,linkcolor=lcolor]{hyperref}

\newcommand{\der}{\mathrm{d}}

\newcommand{\ud}{\, \mathrm{d}}

\newcommand{\tr}{\, \mathrm{Tr} \, }

\newcommand{\nc}{{N_\mathrm{c}}}
\newcommand{\nf}{{n_\mathrm{f}}}

\newcommand{\nr}[1]{(\ref{#1})}

\newcommand{\gev}{\ \textrm{GeV}}

\newcommand{\qs}{Q_\mathrm{s}}

\newcommand{\qso}{Q_\mathrm{s,0}}

\newcommand{\lqcd}{\Lambda_{\mathrm{QCD}}}
\newcommand{\as}{\alpha_{\mathrm{s}}}

\newcommand{\fig}{Fig.~}
\newcommand{\figs}{Figs.~}
\newcommand{\eq}{Eq.~}

\begin{document}

\author{T. Lappi}
\affiliation{
Department of Physics, University of Jyv\"askyl\"a %
 P.O. Box 35, 40014 University of Jyv\"askyl\"a, Finland
}

\affiliation{
Helsinki Institute of Physics, P.O. Box 64, 00014 University of Helsinki,
Finland
}
\author{H. M\"antysaari}
\affiliation{
Department of Physics, University of Jyv\"askyl\"a %
 P.O. Box 35, 40014 University of Jyv\"askyl\"a, Finland
}

\title{
Direct numerical solution of the coordinate space Balitsky-Kovchegov equation at next to leading order
}


\preprint{}

\begin{abstract}
We present the first numerical solution to the next to leading order Balitsky-Kovchegov (BK) equation in coordinate space in the large-$\nc$ limit. In addition to the dipole operator we also solve the evolution of the ``conformal dipole'' for which the conformal invariance breaking double logarithmic term is absent from the evolution equation. The NLO corrections are shown to slow down the evolution. We show that the solution depends strongly on the details of the initial condition, and that the solution to the equation is not positive definite with all initial conditions relevant for phenomenological applications.

\end{abstract}

\maketitle

\section{Introduction}

At the high energies of present day collider experiments, the available phase space for gluon radiation is very large. Since each emitted gluon is itself a source of further emissions, a typical scattering event involves an exponentially growing cascade of gluons. At high enough energy this cascade can fill up the available phase space to the extent that the gluons begin to reinteract. This is the origin of gluon saturation, where the phase space below an energy dependent transverse momentum scale $\qs$ is dominated by nonlinear gluon interactions. This recombination restores the unitarity of the scattering $S$-matrix, which would be violated by an unlimited exponential growth of the cascade.

At high energy, it is convenient to describe QCD scattering off a hadronic target using the eikonal approximation. The natural degrees of freedom describing the gluons in the cascade are then transverse coordinate dependent Wilson lines which describe the eikonal propagation of a high energy quark through the target color field. Cross sections can be expressed in terms of correlators of these Wilson lines. These correlators are universal objects that enable a consistent description of many different scattering phenomena, such as deep inelastic scattering (DIS)~\cite{Albacete:2010sy}, single inclusive particle production in proton-nucleus collisions~\cite{Albacete:2010bs,Lappi:2013zma,Tribedy:2011aa,Rezaeian:2012ye} and two-particle correlations~\cite{Albacete:2010pg,Stasto:2011ru,Lappi:2012nh,
JalilianMarian:2012bd}. In the dilute limit, when nonlinearities are unimportant, these correlators reduce to unintegrated gluon distributions~\cite{Dominguez:2011wm}. This framework is often referred to as the  Color Glass Condensate (CGC, for a review, see e.g. Ref.~\cite{Gelis:2010nm}).

The dependence of these Wilson line correlators on energy, or equivalently Bjorken $x$ or rapidity $y = \ln 1/x$, can be calculated perturbatively even in the nonlinear saturation regime. This energy dependence is described by the Balitsky hierarchy of evolution equations, which for many practical applications can be replaced by its mean field version, the Balitsky-Kovchegov (BK) evolution 
equation~\cite{Balitsky:1995ub,Kovchegov:1999yj}.
It resums large logarithms $\sim\as \ln 1/x$ to all orders. Current phenomenological works typically use the leading order BK equation with the running coupling corrections derived in Ref.~\cite{Balitsky:2006wa}.

While  leading order calculations can give a good physical description of the process, higher order corrections can be numerically very large. It is therefore extremely important for quantitative comparisons with data to  
perform the CGC calculations at next-to-leading (NLO) order accuracy in the QCD coupling $\as$. First steps in this direction have been taken in particle production in pA collisions~\cite{Chirilli:2011km,Chirilli:2012jd,Stasto:2013cha,Altinoluk:2014eka} and DIS~\cite{Balitsky:2010ze,Beuf:2011xd}.

A crucial ingredient of a consistent NLO treatment in this context is solving the NLO BK equation, which describes the energy or rapidity, dependence of the Wilson line correlators. This equation has been derived in Ref.~\cite{Balitsky:2008zza}. Its linearized limit, the NLO~BFKL equation has been known already before~\cite{Fadin:1996nw,Fadin:1998py,Ciafaloni:1998gs}. It has been noted that it is affected with contributions from large transverse momentum logarithms, a feature that seems to remain true when the NLO~BFKL equation is complemented with an absorptive boundary condition to emulate the nonlinear effects~\cite{Avsar:2011ds}.
Elaborate resummation schemes have been proposed to treat these contributions~\cite{Salam:1998tj,Ciafaloni:1999yw,Altarelli:1999vw,Ciafaloni:2003rd}. These resummations, however, rely on the linear structure of the equation and cannot easily be generalized to BK. 
There have also been proposals for a kinematically constrained version of the BK equation~\cite{Motyka:2009gi,Beuf:2014uia} to solve these issues.
While it is expected that these same instabilities also manifest themselves in the  fully nonlinear BK equation in coordinate space, this has never been shown explicitly. In order to directly address this question, we will in this work solve numerically  the next to leading order BK evolution equation in the form derived in Ref.~\cite{Balitsky:2008zza}.
We will also study numerically the equation for the ``composite conformal dipole'',
proposed  in Ref.~\cite{Balitsky:2009xg} to address some of the issues with the NLO~BK equation. The precise forms of the equations studied numerically are written down in Secs.~\ref{sec:nlobk} and~\ref{sec:confnlobk}. Our numerical results are then discussed 
in Sec.~\ref{sec:numerics}. In principle the dipole scattering amplitude obtained here could be convoluted with the NLO photon impact factor from \cite{Balitsky:2010ze,Beuf:2011xd} for a comparison with experimental data. We will demonstrate, however, that this would be difficult since the equation itself is unstable for many values of the initial conditions in the phenomenologically relevant regime.

\section{Balitsky-Kovchegov equation at NLO}
\label{sec:nlobk}

We study in this work the next-to-leading order BK evolution equation derived in~\cite{Balitsky:2008zza}, which we write as:
\begin{multline}
\label{eq:nlobk}
	\partial_y S(r) = \frac{\as \nc}{2\pi^2} K_1 \otimes [S(X)S(Y)-S(r)] \\
		+ \frac{\as^2 \nc^2}{8\pi^4} K_2 \otimes [S(X)S(z-z')S(Y')-S(X)S(Y)] \\
		+ \frac{\as^2 \nf \nc}{8\pi^4} K_f \otimes S(Y)[S(X')-S(X)]
\end{multline}
Here we have taken the large-$\nc$ and  mean field limits to express the equation in a closed form in terms of only the correlator of two Wilson lines (the ``dipole''):
\begin{equation}
	\label{eq:s-def}
	S(r) = \frac{1}{\nc} \langle \tr U_x U^\dagger_y\rangle.
\end{equation}
Here the brackets $\langle \rangle$ stand for an average over the target color field.
The kernels appearing in \eq \eqref{eq:nlobk} are
\begin{widetext}
\begin{align}
K_1 &= \frac{r^2}{X^2Y^2} \left[ 1+\frac{\as\nc }{4\pi} \left(  \frac{\beta}{\nc} \ln r^2\mu^2 - \frac{\beta}{\nc} \frac{X^2-Y^2}{r^2} \ln \frac{X^2}{Y^2} + \frac{67}{9} - \frac{\pi^2}{3} - \frac{10}{9} \frac{\nf}{\nc} - \ln \frac{X^2}{r^2} \ln \frac{Y^2}{r^2} \right) \right]
\\
K_2 &= -\frac{2}{(z-z')^4} + \left[ \frac{X^2 Y'^2 + X'^2Y^2 - 4r^2(z-z')^2}{(z-z')^4(X^2Y'^2 - X'^2Y^2)} + \frac{r^4}{X^2Y'^2(X^2Y'^2 - X'^2Y^2)} + \frac{r^2}{X^2Y'^2(z-z')^2} \right]\\
&  \times  \ln \frac{X^2Y'^2}{X'^2Y^2} \nonumber
\\
 K_f &= \frac{2}{(z-z')^4}  
	- \frac{X'^2Y^2 + Y'^2 X^2 - r^2 (z-z')^2}{(z-z')^4(X^2Y'^2 - X'^2Y^2)} \ln \frac{X^2Y'^2}{X'^2Y^2} 
\end{align}
\end{widetext}
The coordinates 
are two dimensional vectors denoted as $r=x-y$, $X=x-z$, $Y=y-z$, $X'=x-z'$ and $Y'=y-z'$. The convolutions $\otimes$ are calculated by integrating over the vectors $z$ and $z'$. The kernel $K_1$ consists of the leading order BK kernel $r^2/(X^2Y^2)$ and an NLO correction $\sim \as$, and the beta function coefficient is $\beta = \frac{11}{3}\nc - \frac{2}{3}\nf$ with $\nf=3$. 

Part of the NLO corrections, especially the term involving the renormalization scale $\mu^2$, should be absorbed into the running of the strong coupling $\as$. What terms exactly are  absorbed in $\as$, and at which scale it is evaluated, is a scheme choice. We adapt the choice derived in Ref.~\cite{Balitsky:2006wa} and replace all terms in $K_1$ proportional to the $\beta$ function by the Balitsky running coupling prescription that is also used to solve the leading order BK equation with running coupling corrections. For the other terms we choose to evaluate $\as$ at the scale given by the size of the parent dipole $r$. We thus write the first kernel as
\begin{widetext}
\begin{multline}
	\frac{\as \nc}{2\pi^2} K_1 = \frac{\as(r) \nc}{2\pi^2} \left[\frac{r^2}{X^2Y^2} + \frac{1}{X^2} \left(\frac{\as(X)}{\as(Y)}-1\right) + \frac{1}{Y^2} \left(\frac{\as(Y)}{\as(X)}-1\right) \right] \\
		+ \frac{\as(r)^2 \nc^2}{8\pi^3} \frac{r^2}{X^2Y^2} \left[ \frac{67}{9} - \frac{\pi^2}{3} - \frac{10}{9} \frac{\nf}{\nc} - 2\ln \frac{X^2}{r^2} \ln \frac{Y^2}{r^2} \right]
\end{multline}
\end{widetext}

We use the same expression of the coupling in terms of $r^2$ as in Ref.~\cite{Lappi:2012vw}.
While the equation is simpler to write in terms of $S$, the results of our calculation will be expressed in terms of the scattering amplitude\footnote{Note that a term $N(Y') - N(Y)$ is missing from Eq. (136) of Ref.~\cite{Balitsky:2008zza}} $N=1-S$.

At finite $\nc$, correlators of  up to six Wilson lines would also be needed in order to evaluate the rapidity derivative of the dipole operator in  \eq\eqref{eq:nlobk}. In principle one could obtain the higher-point functions from a solution of the NLO JIMWLK equation~\cite{Balitsky:2013fea,Kovner:2013ona}, or using e.g. a Gaussian approximation which allows one to write any higher point function in terms of the dipole operators as described in Ref.~\cite{Dominguez:2012ad}.
The contribution from finite-$\nc$ corrections to the leading order BK equation have been studied in Ref.~\cite{Kovchegov:2008mk}, and their contribution is numerically found to be even smaller than the $\sim 1/\nc^2$ one would naively expect, so we feel justified in neglecting them here.

\begin{figure}[tb]
\begin{center}
\includegraphics[width=0.49\textwidth]{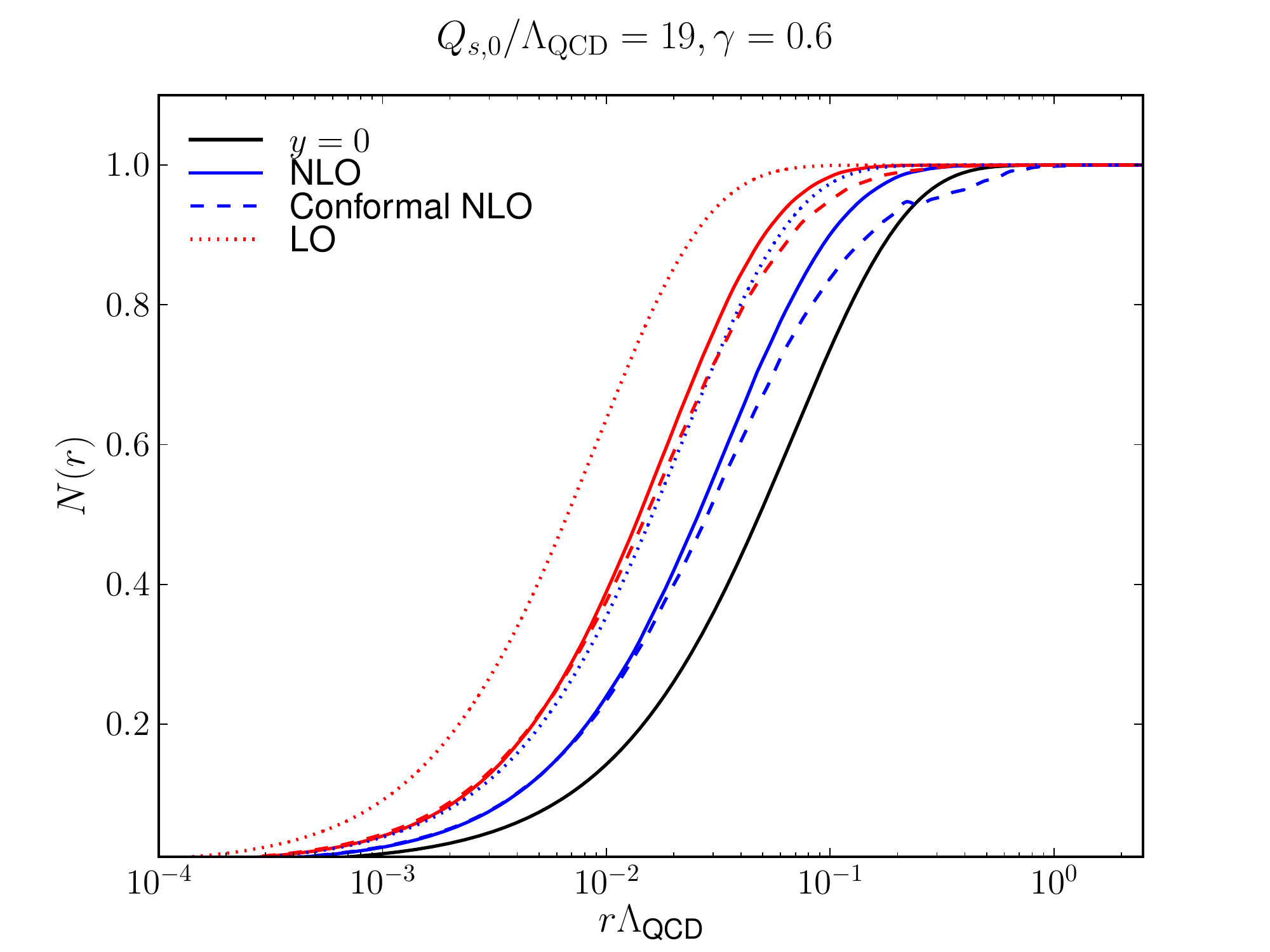}
\end{center}
\caption{
Dipole amplitude and conformal dipole amplitude at initial condition and after evolution compared to the solution of the LO BK equation at rapidities $y=0, 5$ and $y=10$ (from right to left).
}\label{fig:amplitude}
\end{figure}

\begin{figure}[tb]
\begin{center}
\includegraphics[width=0.49\textwidth]{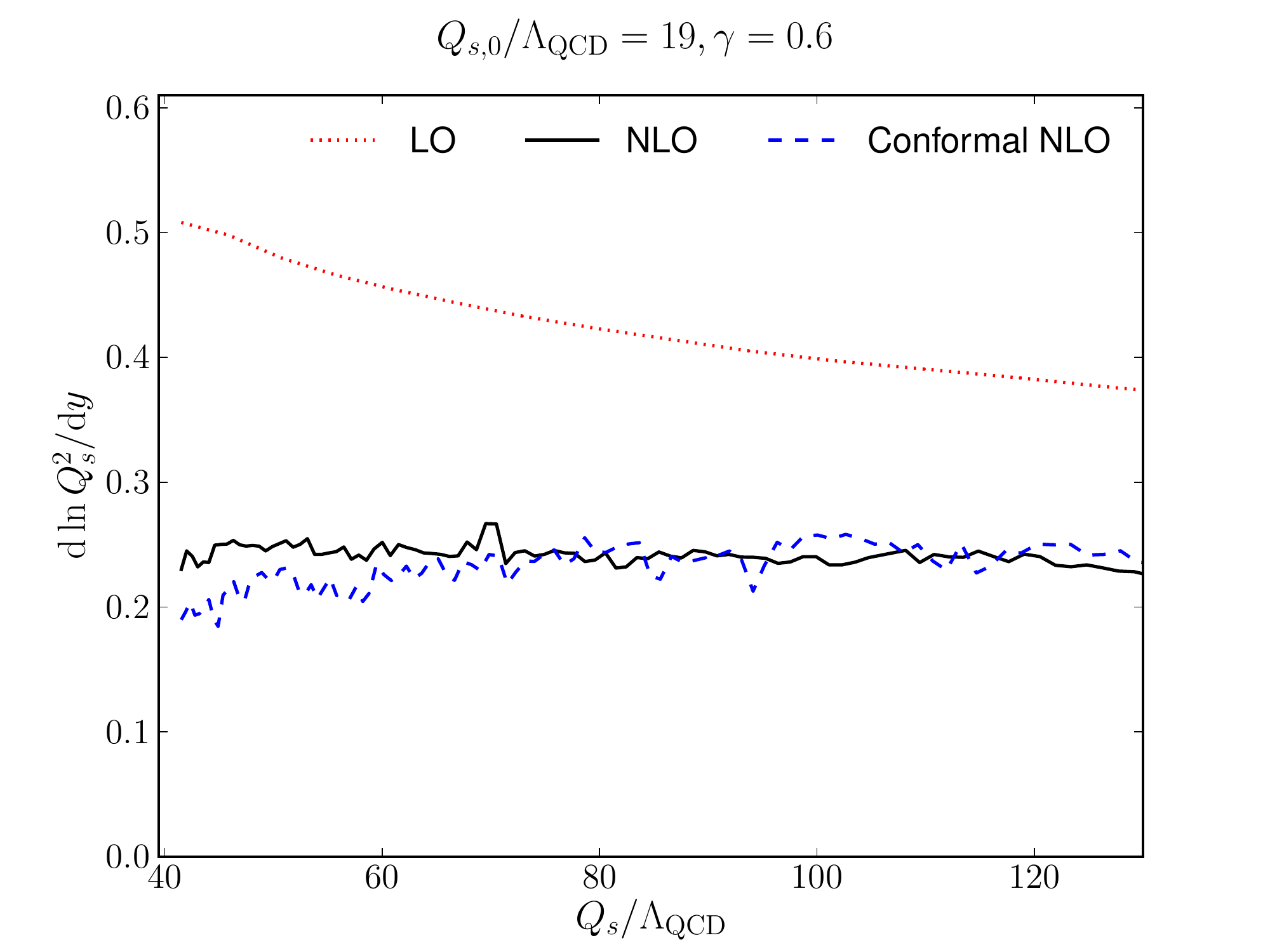}
\end{center}
\caption{
Evolution speed for the conformal and non-conformal dipoles as a a function of the saturation scale compared to the leading order BK equation solution. 
}\label{fig:dqsdy}
\end{figure}

\begin{figure*}[ptb]
	\subfloat[$\gamma=0.6$]{
		\includegraphics[width=0.33\textwidth]{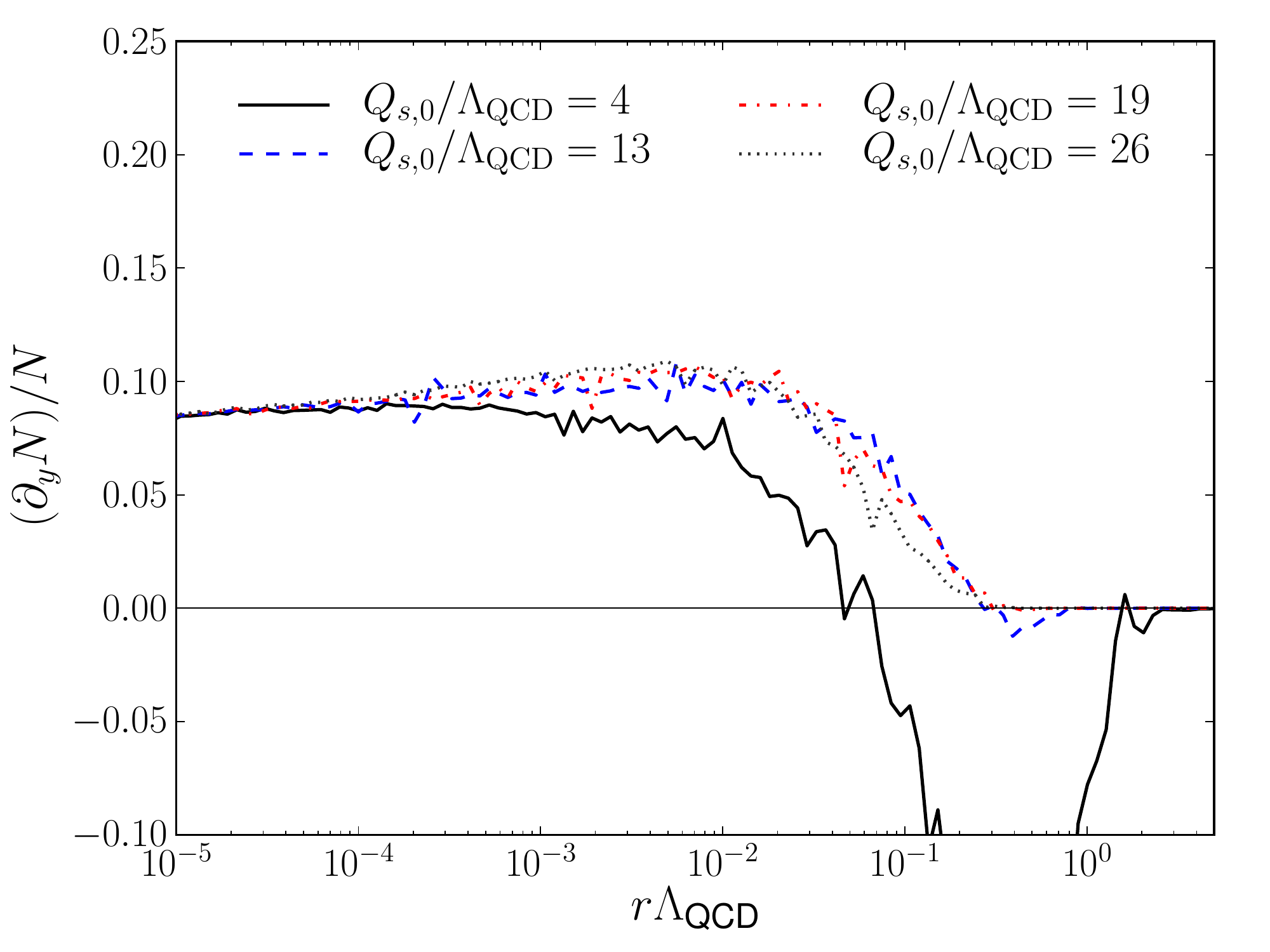}
		}
	\subfloat[$\gamma=0.8$]{
		\includegraphics[width=0.33\textwidth]{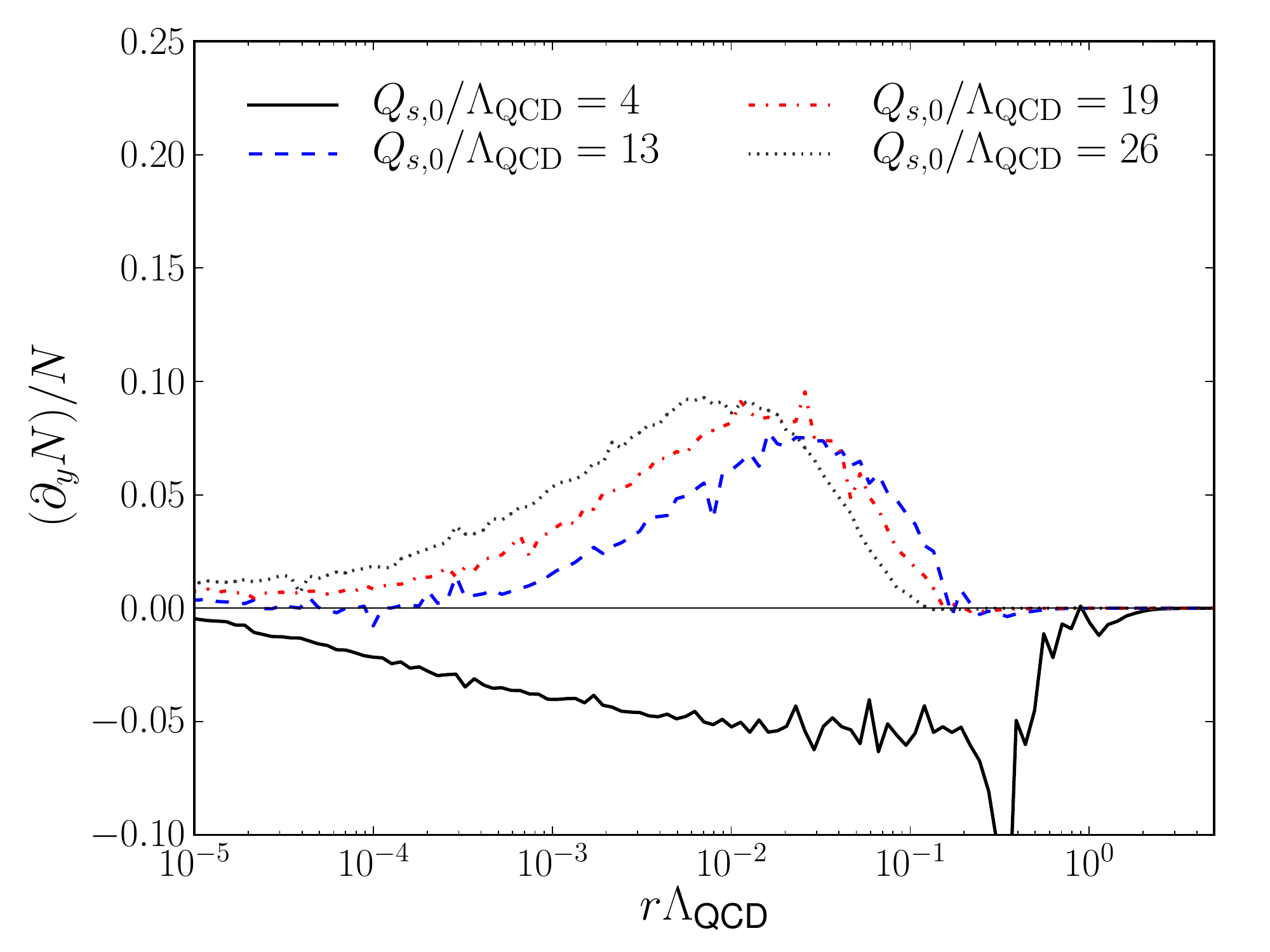}
		}
	\subfloat[$\gamma=1.0$]{
		\includegraphics[width=0.33\textwidth]{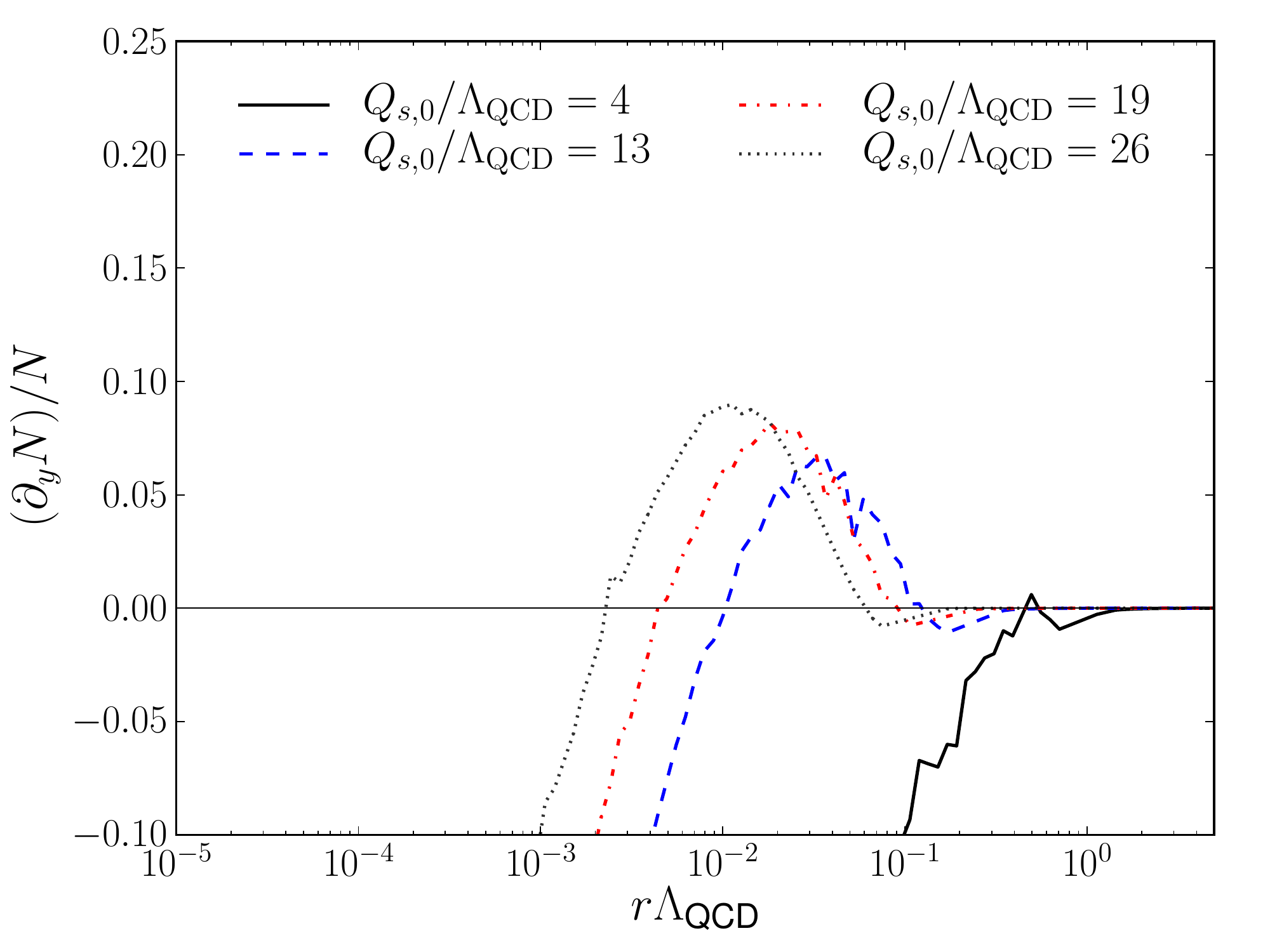}
		}	
	\caption{Logarithmic derivative of the dipole amplitude (evolution speed) at initial condition with different values for the anomalous dimension.}
	\label{fig:dndy_dipole_y0}
\end{figure*}

\begin{figure*}[ptb]
	\subfloat[$\gamma=0.6$]{
		\includegraphics[width=0.33\textwidth]{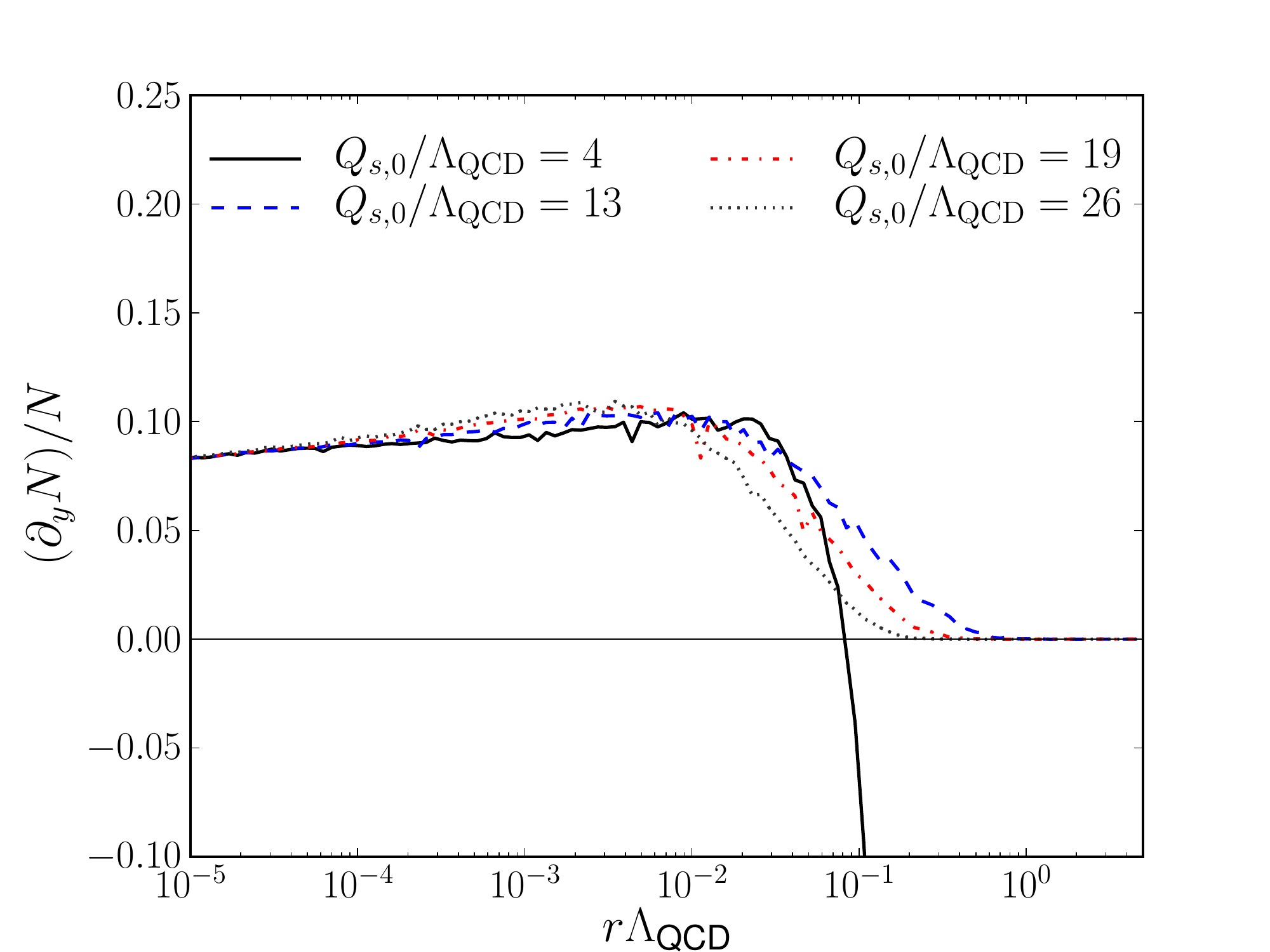}
		}
	\subfloat[$\gamma=0.8$]{
		\includegraphics[width=0.33\textwidth]{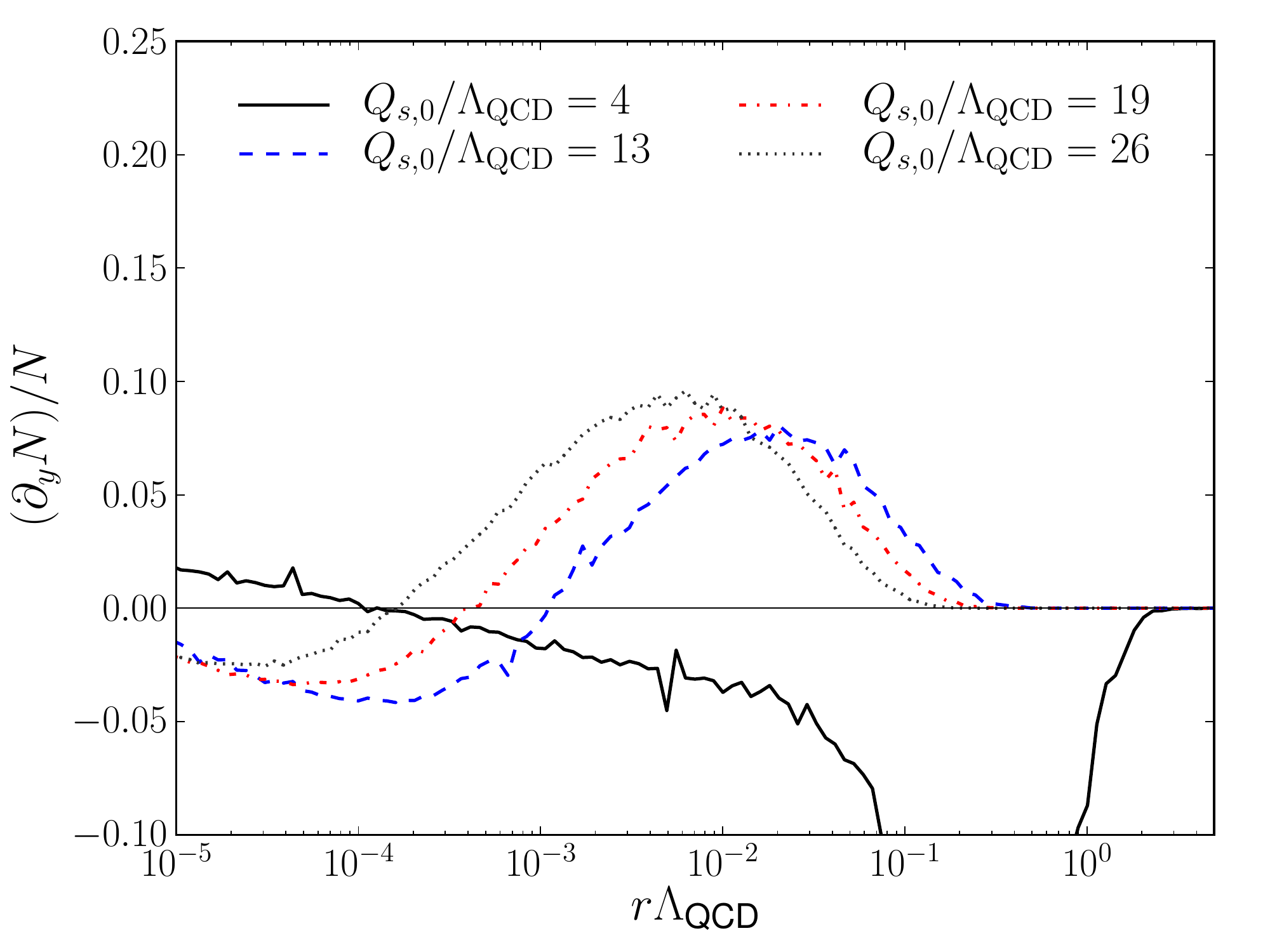}
		}
	\subfloat[$\gamma=1.0$]{
		\includegraphics[width=0.33\textwidth]{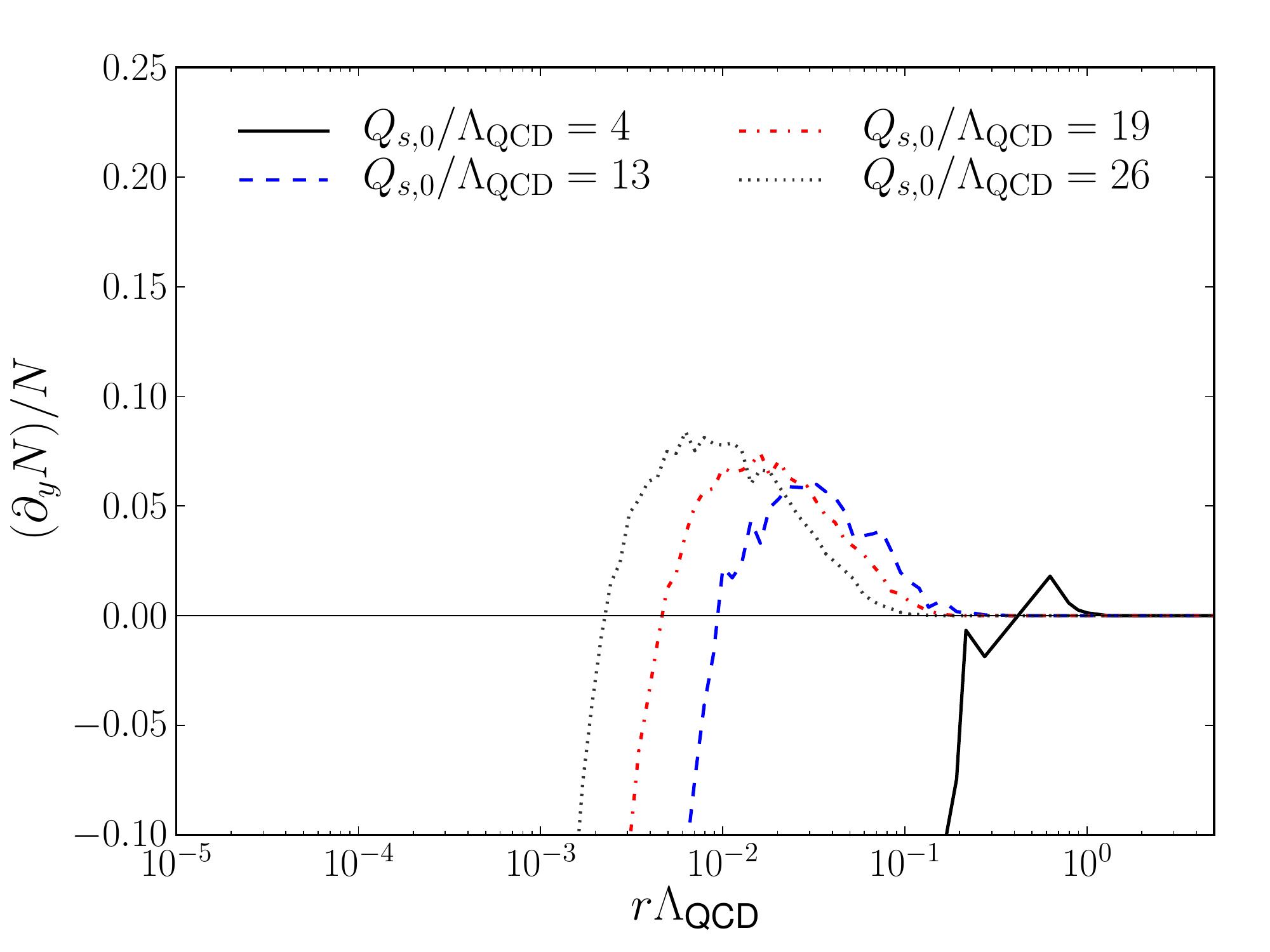}
		}	
	\caption{Evolution speed of the dipole amplitude at $y=5$ with different values for the anomalous dimension at the initial condition.}
	\label{fig:dndy_dipole_y5}
\end{figure*}

\section{Evolution equation for the conformal dipole}
\label{sec:confnlobk}

The Wilson lines are conformally invariant, and thus their evolution equation should be conformal in a conformal field theory. In QCD, one expects the conformal invariance to be broken only by the running of the coupling. However, the evolution \eq\eqref{eq:nlobk} also has a conformal symmetry breaking NLO double logarithmic term $\ln X^2/r^2 \ln Y^2/r^2$ in the kernel $K_1$. 
Diagrammatically this contribution arises from the diagrams with  a loop in the  $1\to 2$ dipole transition where one gluon interacts with the shockwave, see discussion and Fig. 9 in Ref.~\cite{Balitsky:2008zza}.
The reason for the conformal invariance breaking is the fact that the derivation of Ref.~\cite{Balitsky:2008zza} uses a cutoff in the longitudinal direction that violates the symmetry. This was confirmed by the appearence of the same nonconformal double logarithm in the fully conformally invariant $N=4$ supersymmetric Yang-Mills (SYM) theory \cite{Balitsky:2009xg}.

A possible way to restore the conformal invariance, proposed in Ref.~\cite{Balitsky:2009xg}, is to rewrite the evolution equation in terms of the  conformal dipole $S^\text{conf}$, defined as
\begin{multline}
\label{eq:conf-n}
S(r)^\text{conf} = S(r) \\- \frac{\as \nc}{4\pi^2} \int \der^2 z \frac{r^2}{X^2Y^2} \ln \frac{a r^2}{X^2Y^2} [ S(X)S(Y) - S(r) ].
\end{multline}
Here $a$ is an arbitrary dimensional constant which will eventually cancel from the evolution equation.
Using \eq\eqref{eq:nlobk} one can then derive the NLO evolution equation for the conformal dipole. The resulting equation turns out to differ from the NLO BK equation only by  
the  disappearance of the double logarithmic term $\ln X^2/r^2 \ln Y^2/r^2$ from $K_1$, and the appearance of an additional contribution 
\begin{equation}
	\frac{2r^2}{X^2Y'^2(z-z')^2} \ln \frac{r^2(z-z')^2}{X'^2Y^2}
\end{equation}
in the kernel $K_2$.
 Now the only term that breaks the conformal invariance is the running coupling $\as$. 
The corresponding evolution equation in $N=4$ SYM theory is fully conformal~\cite{Balitsky:2009xg}.

\section{Numerical analysis}
\label{sec:numerics}

\begin{figure*}[ptb]
	\subfloat[$\gamma=0.6$]{
		\includegraphics[width=0.33\textwidth]{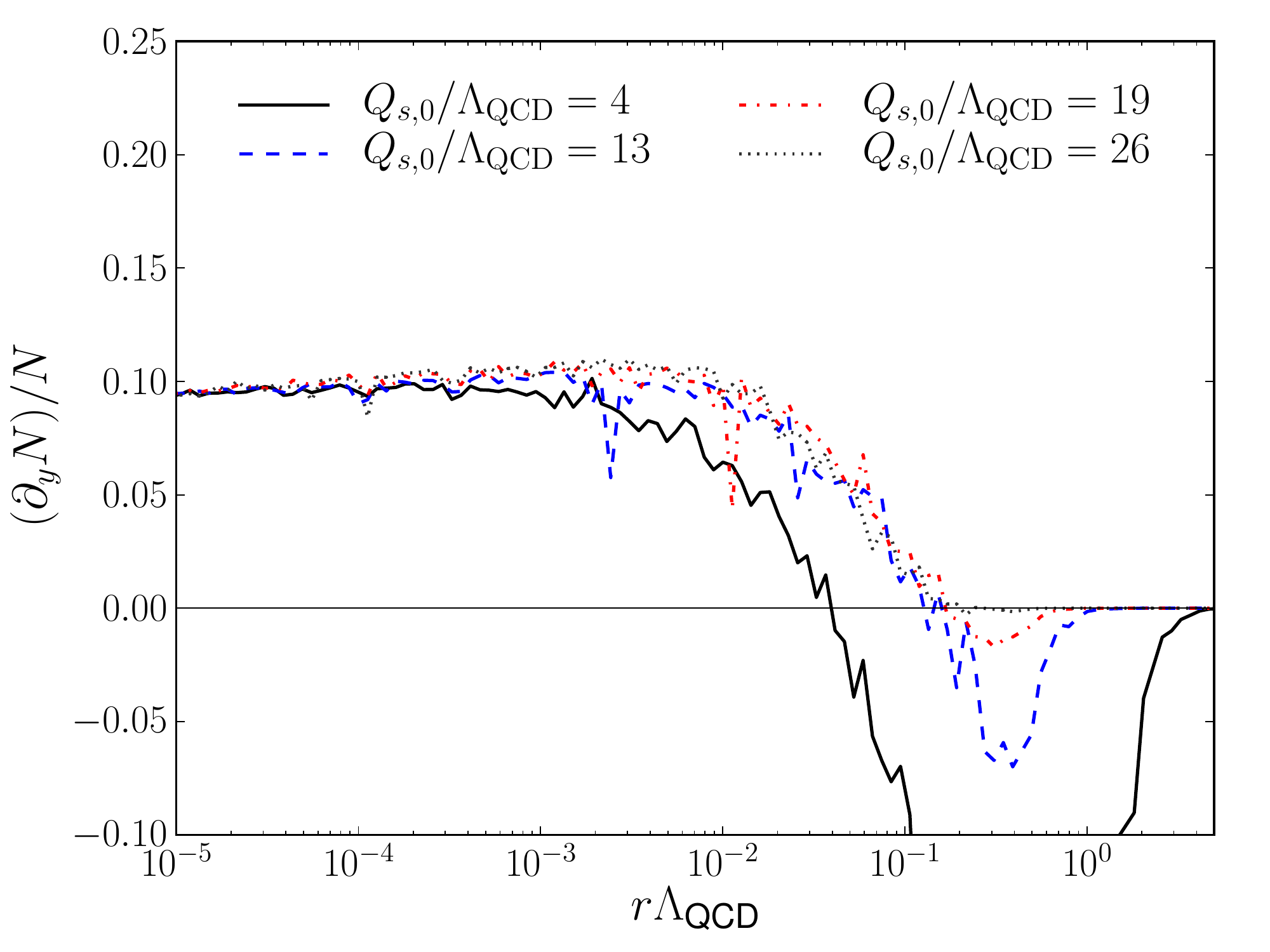}
		}
	\subfloat[$\gamma=0.8$]{
		\includegraphics[width=0.33\textwidth]{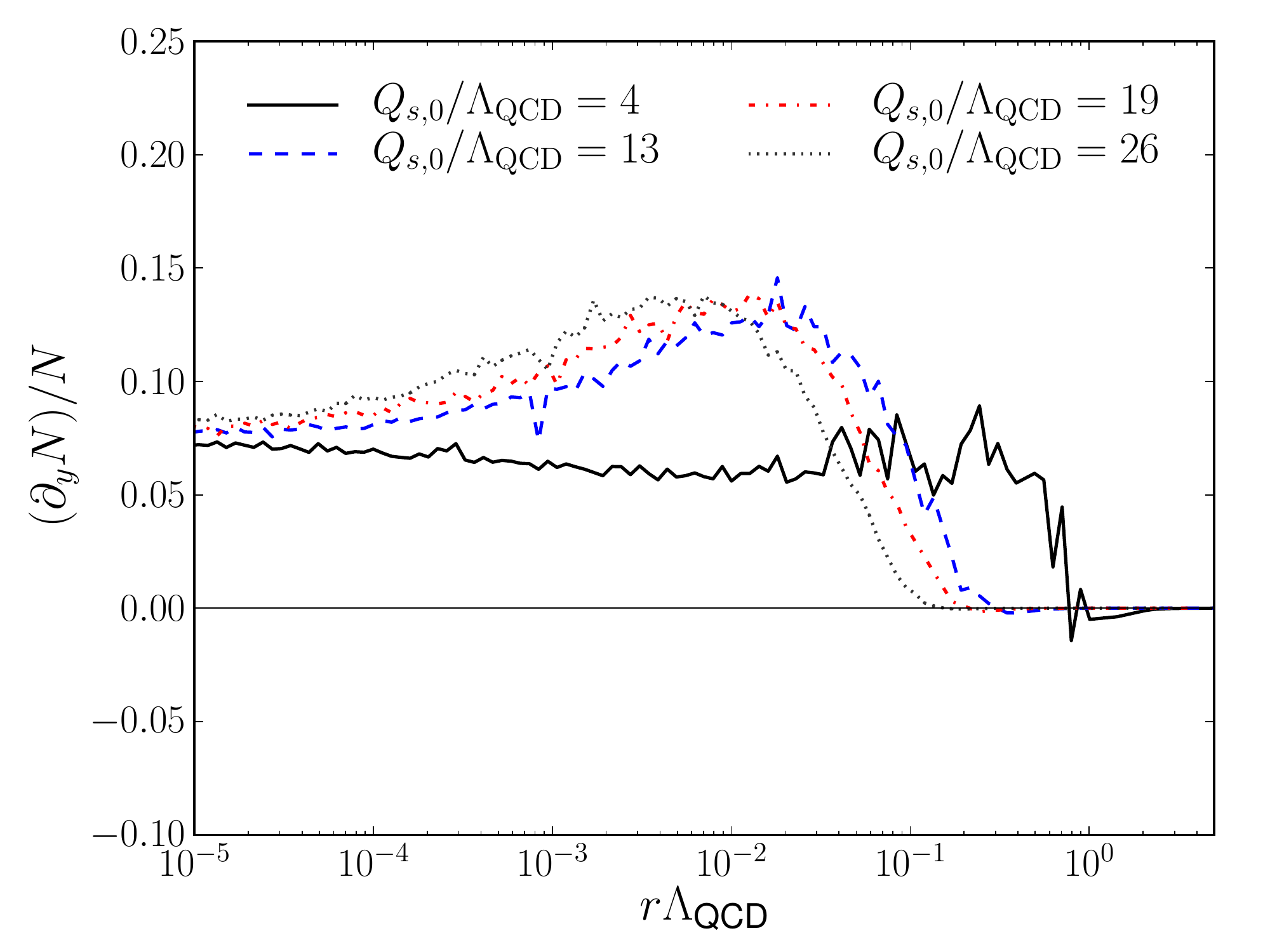}
		}
	\subfloat[$\gamma=1.0$]{
		\includegraphics[width=0.33\textwidth]{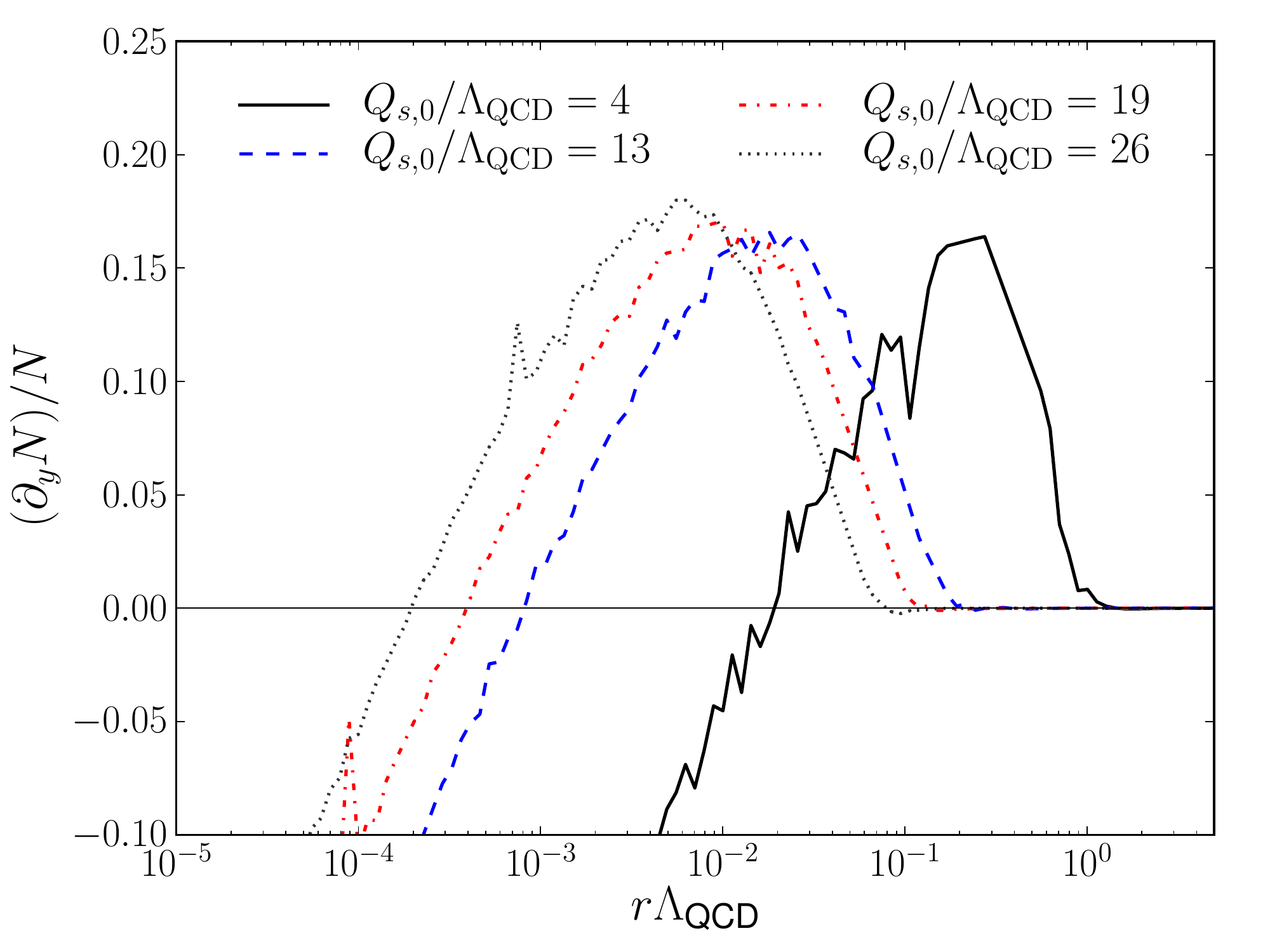}
		}	
	\caption{Evolution speed of the conformal dipole amplitude at initial condition with different values for the anomalous dimension.}
	\label{fig:dndy_confdipole_y0}
\end{figure*}

\begin{figure*}[ptb]
	\subfloat[$\gamma=0.6$]{
		\includegraphics[width=0.33\textwidth]{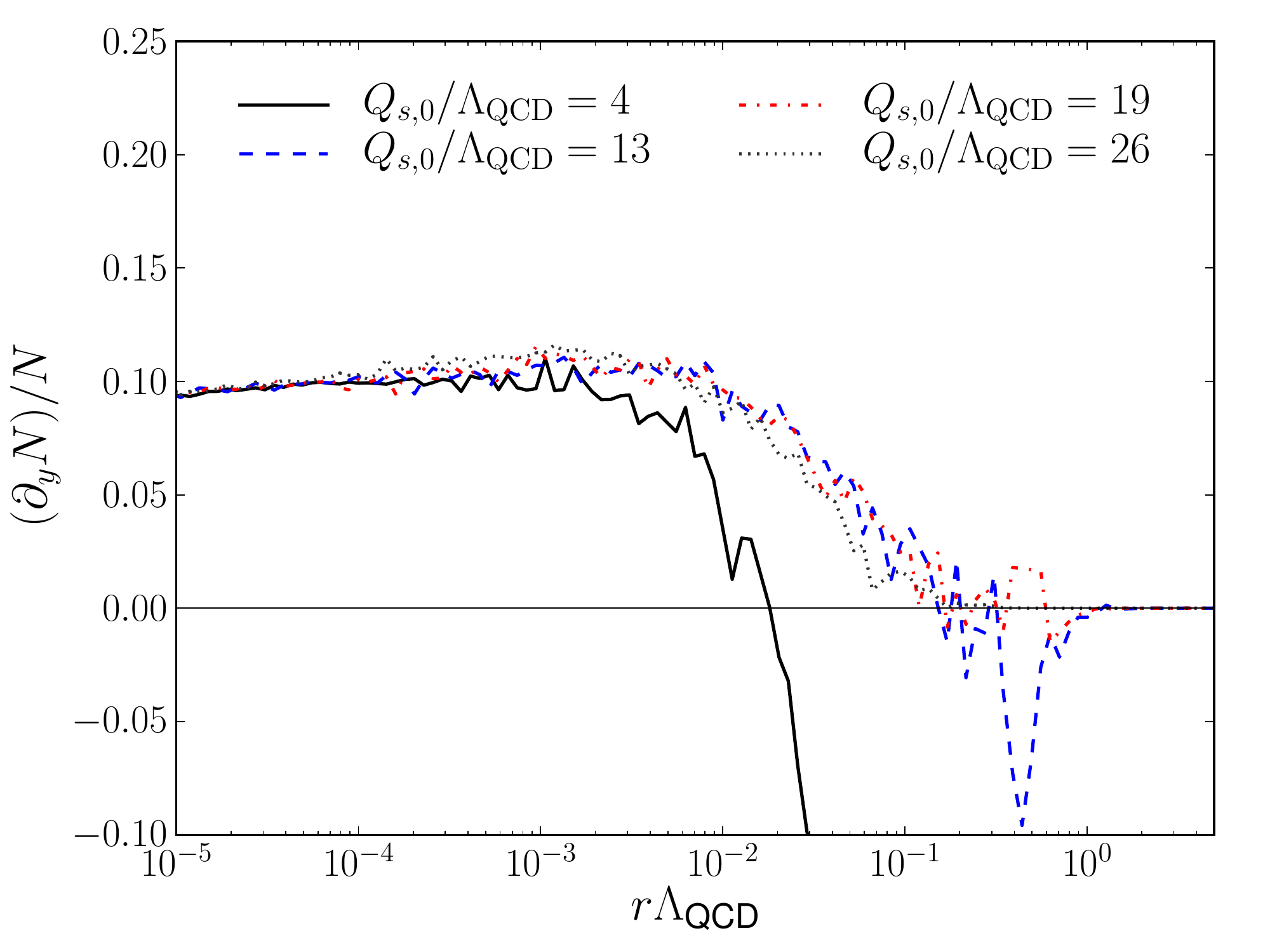}
		}
	\subfloat[$\gamma=0.8$]{
		\includegraphics[width=0.33\textwidth]{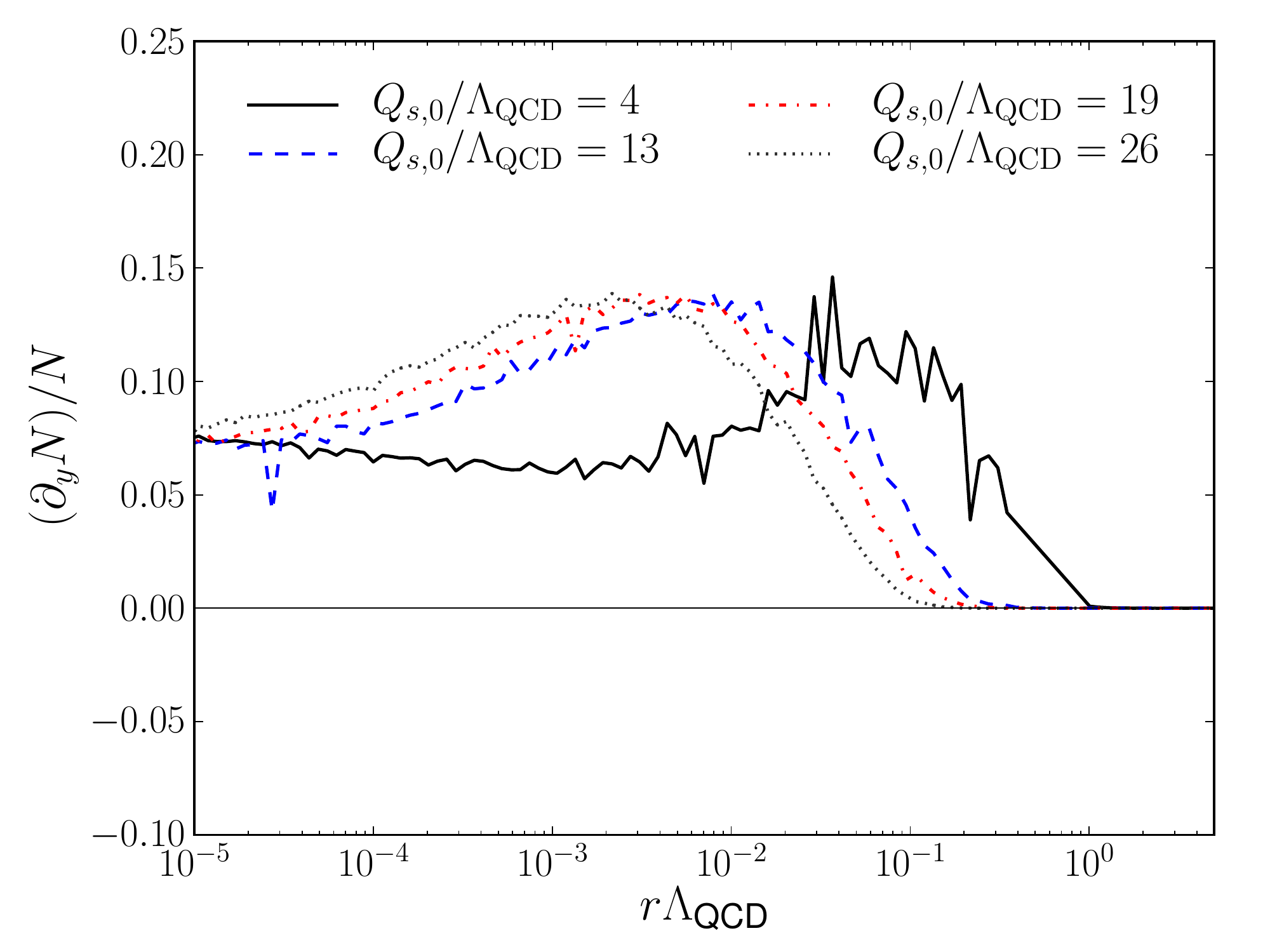}
		}
	\subfloat[$\gamma=1.0$]{
		\includegraphics[width=0.33\textwidth]{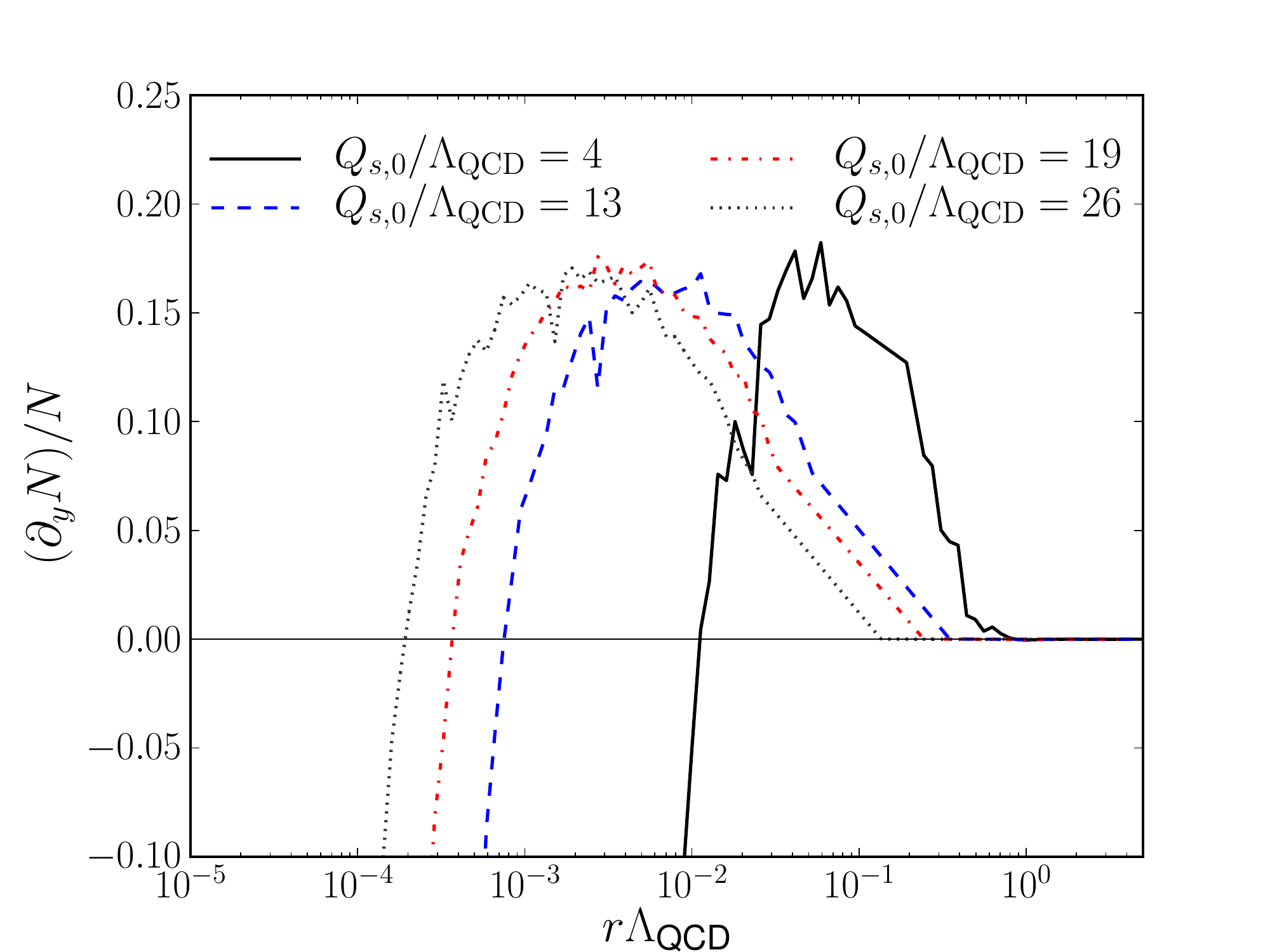}
		}	
	\caption{Evolution speed of the conformal dipole amplitude at $y=5$ with different values for the anomalous dimension at the initial condition.}
	\label{fig:dndy_confdipole_y5}
\end{figure*}

\begin{figure*}[ptb]
	\subfloat[$y=1$]{
		\includegraphics[width=0.33\textwidth]{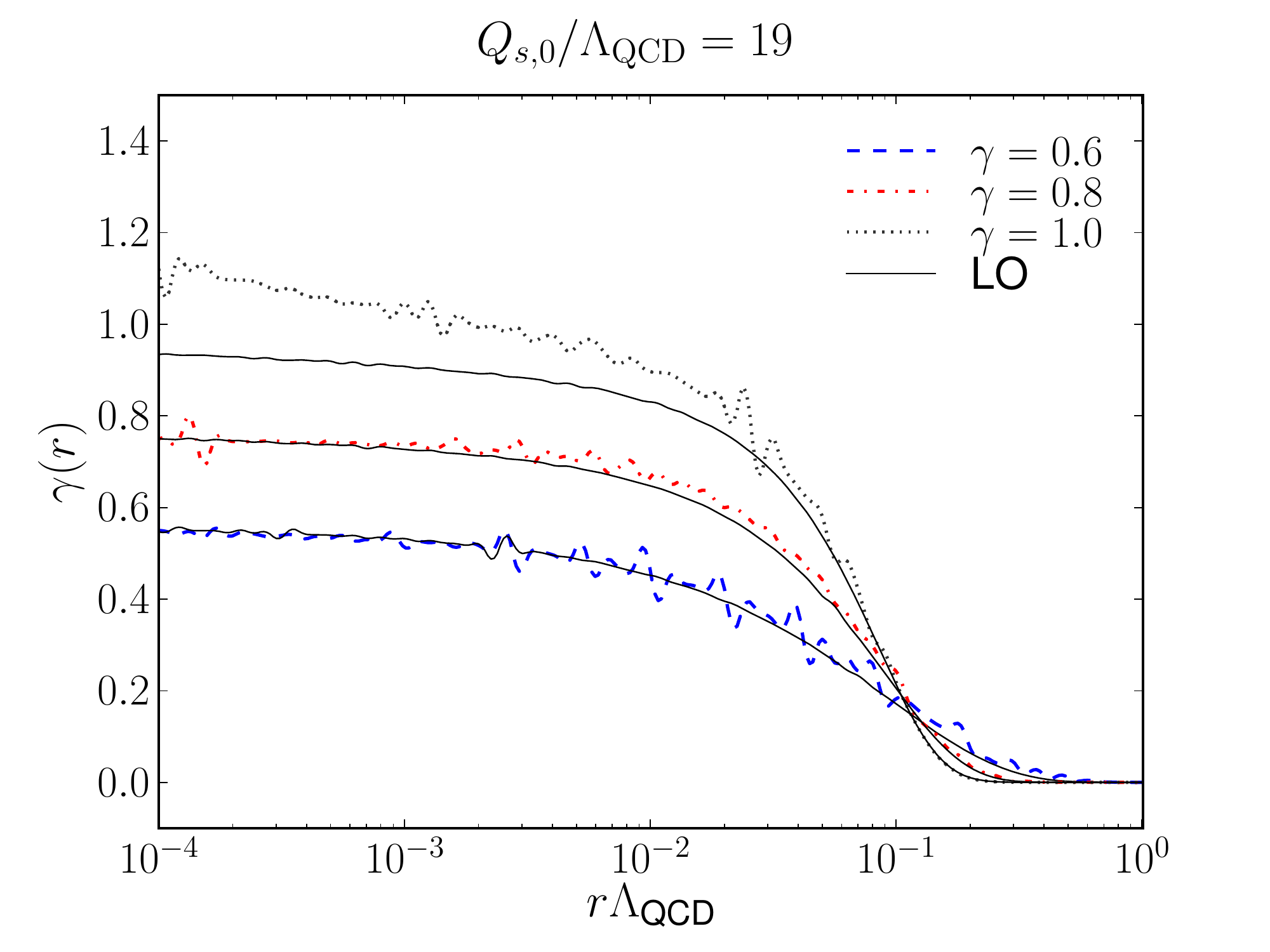}
		}
	\subfloat[$y=5$]{
		\includegraphics[width=0.33\textwidth]{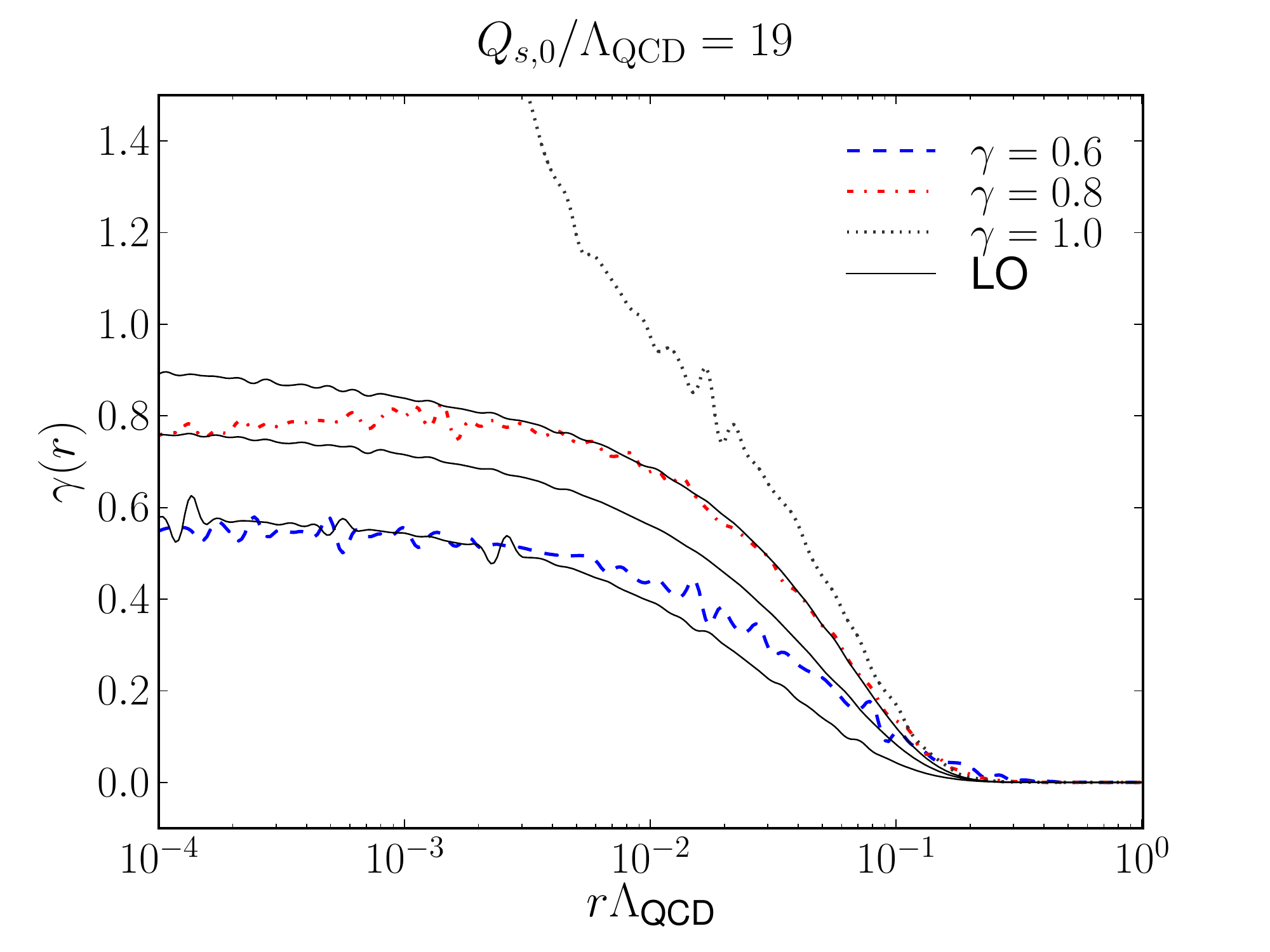}
		}
	\subfloat[$y=30$]{
		\includegraphics[width=0.33\textwidth]{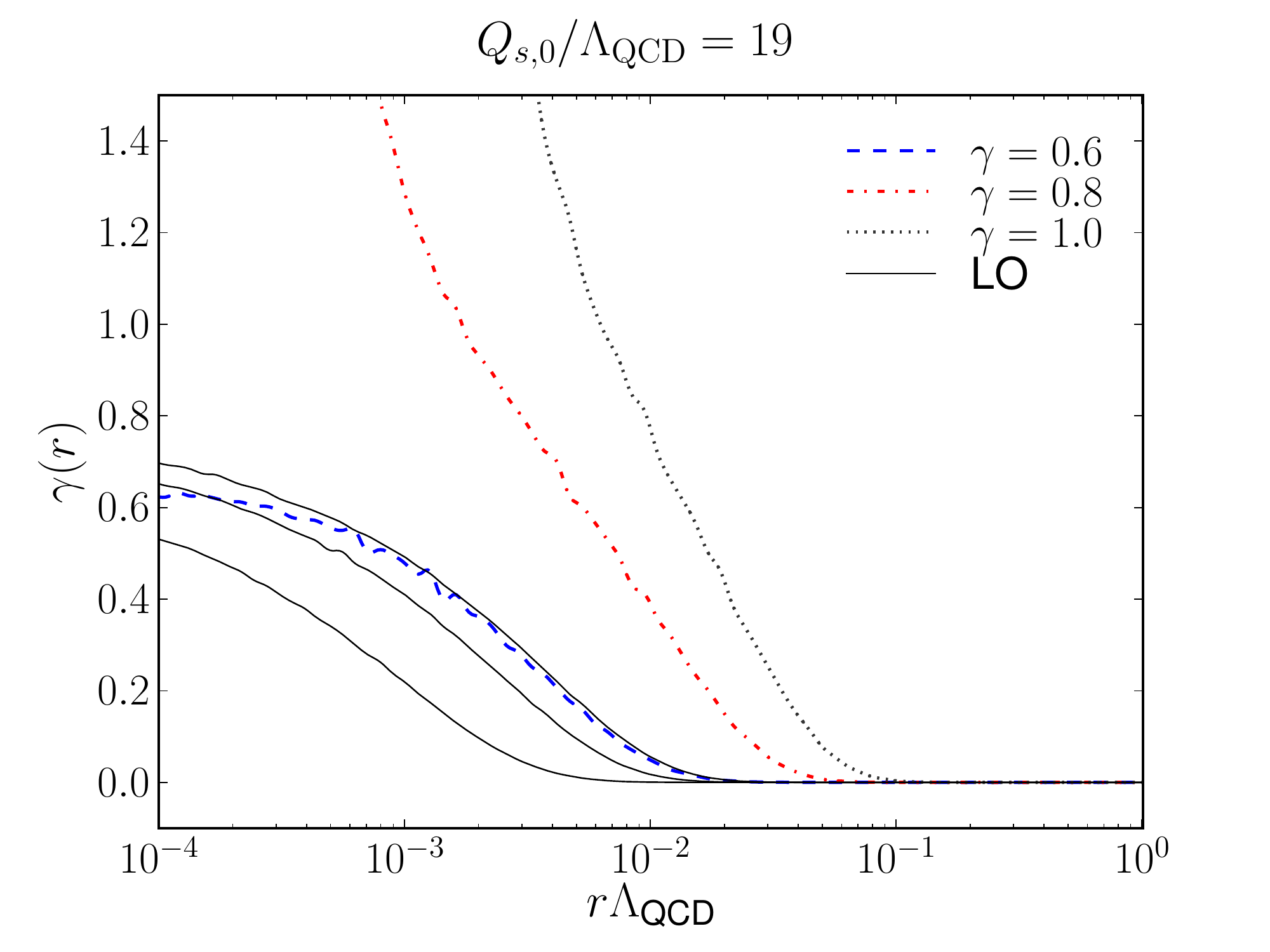}
		}			
	\caption{Anomalous dimension $\gamma(r)=\der \ln N(r)/\der \ln r^2$ as a function of dipole size at different rapidities. The plots from right to left are for different rapidities  $y=1,5,30$. Shown in each plot are the solutions to the evolution equation with different initial anomalous dimensions. The solid lines are the solutions of the LO equation and the dashed lines the non-conformal dipoles, with different dashed lines corresponding to different initial anomalous dimensions.}
	\label{fig:gammaqcd}
\end{figure*}

\begin{figure*}[ptb]
	\subfloat[$y=1$]{
		\includegraphics[width=0.33\textwidth]{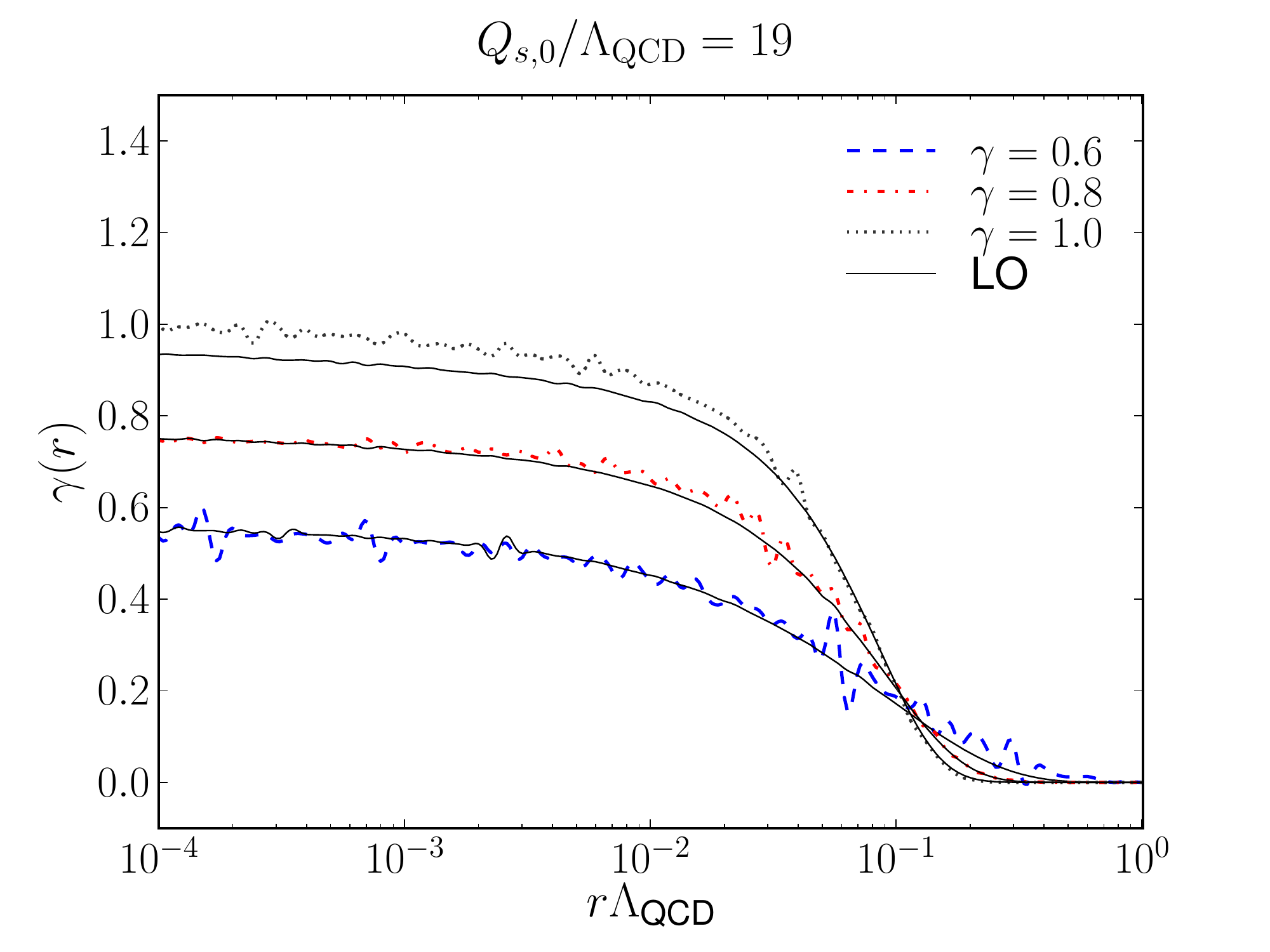}
		}
	\subfloat[$y=5$]{
		\includegraphics[width=0.33\textwidth]{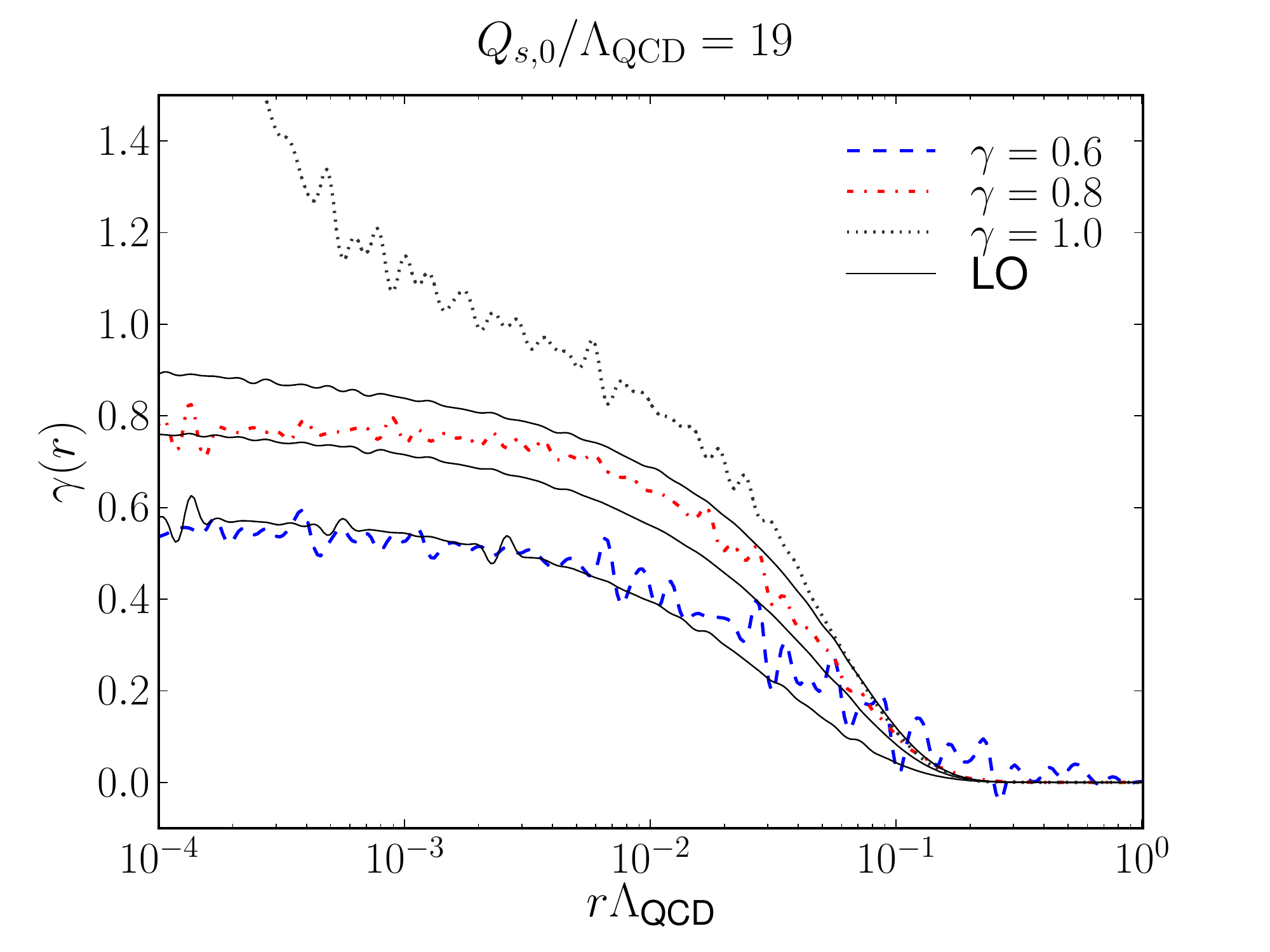}
		}
	\subfloat[$y=30$]{
		\includegraphics[width=0.33\textwidth]{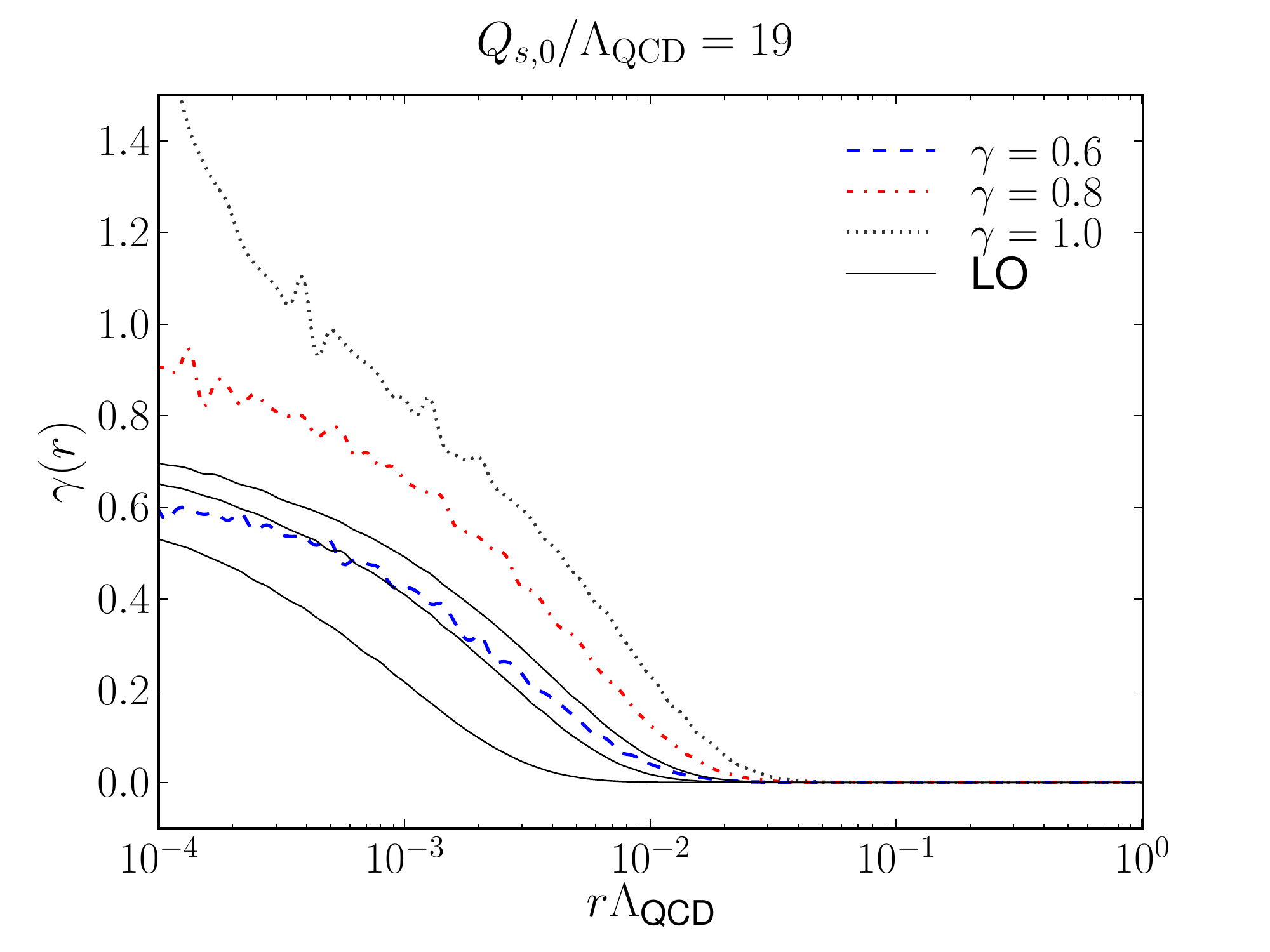}
		}			
	\caption{Same as Fig.~\ref{fig:gammaqcd} for the conformal dipoles.}
	\label{fig:gammaconfqcd}
\end{figure*}

\begin{figure*}[ptb]
	\subfloat[$\gamma=0.6$]{
		\includegraphics[width=0.33\textwidth]{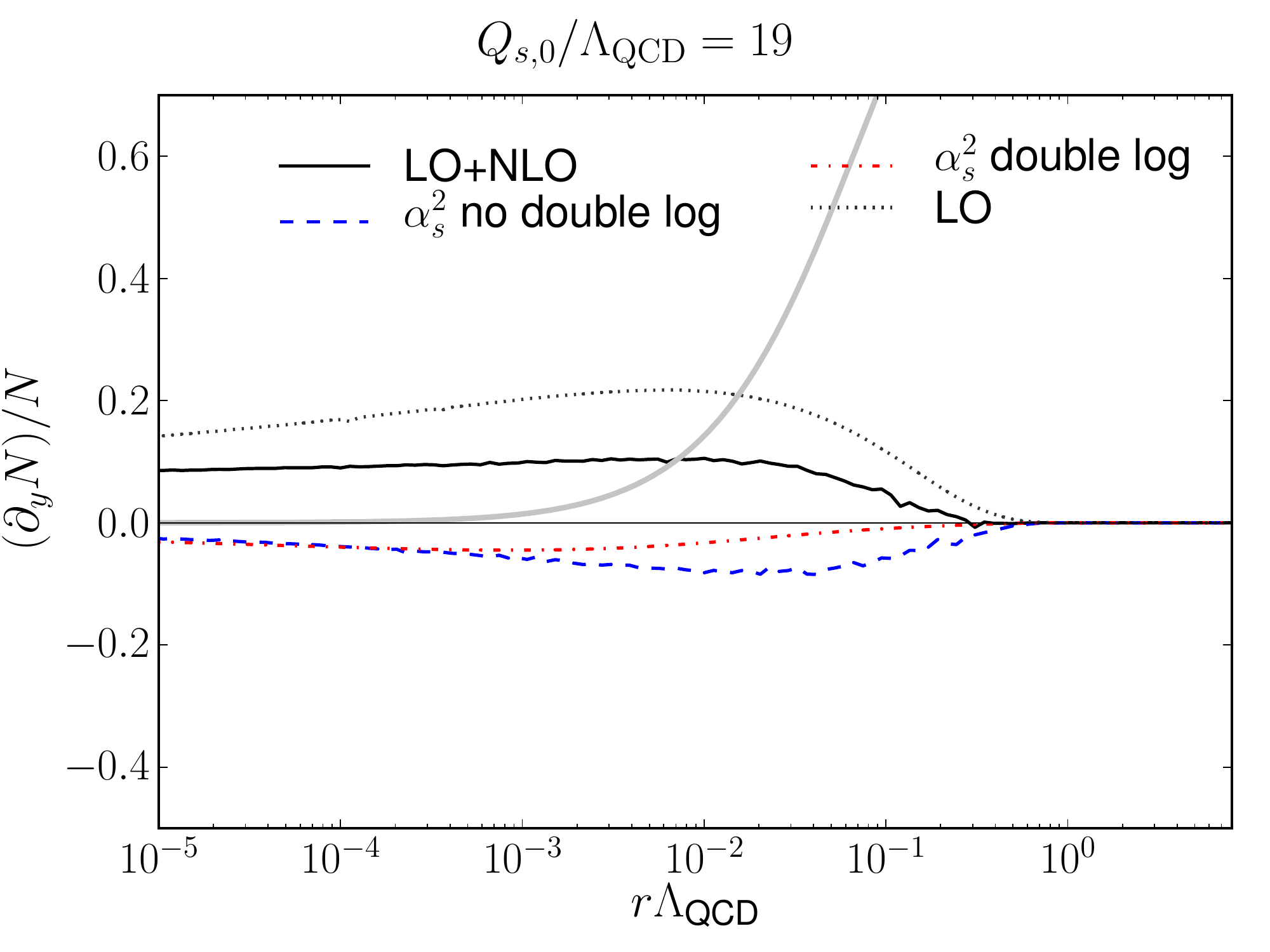}
		}
	\subfloat[$\gamma=0.8$]{
		\includegraphics[width=0.33\textwidth]{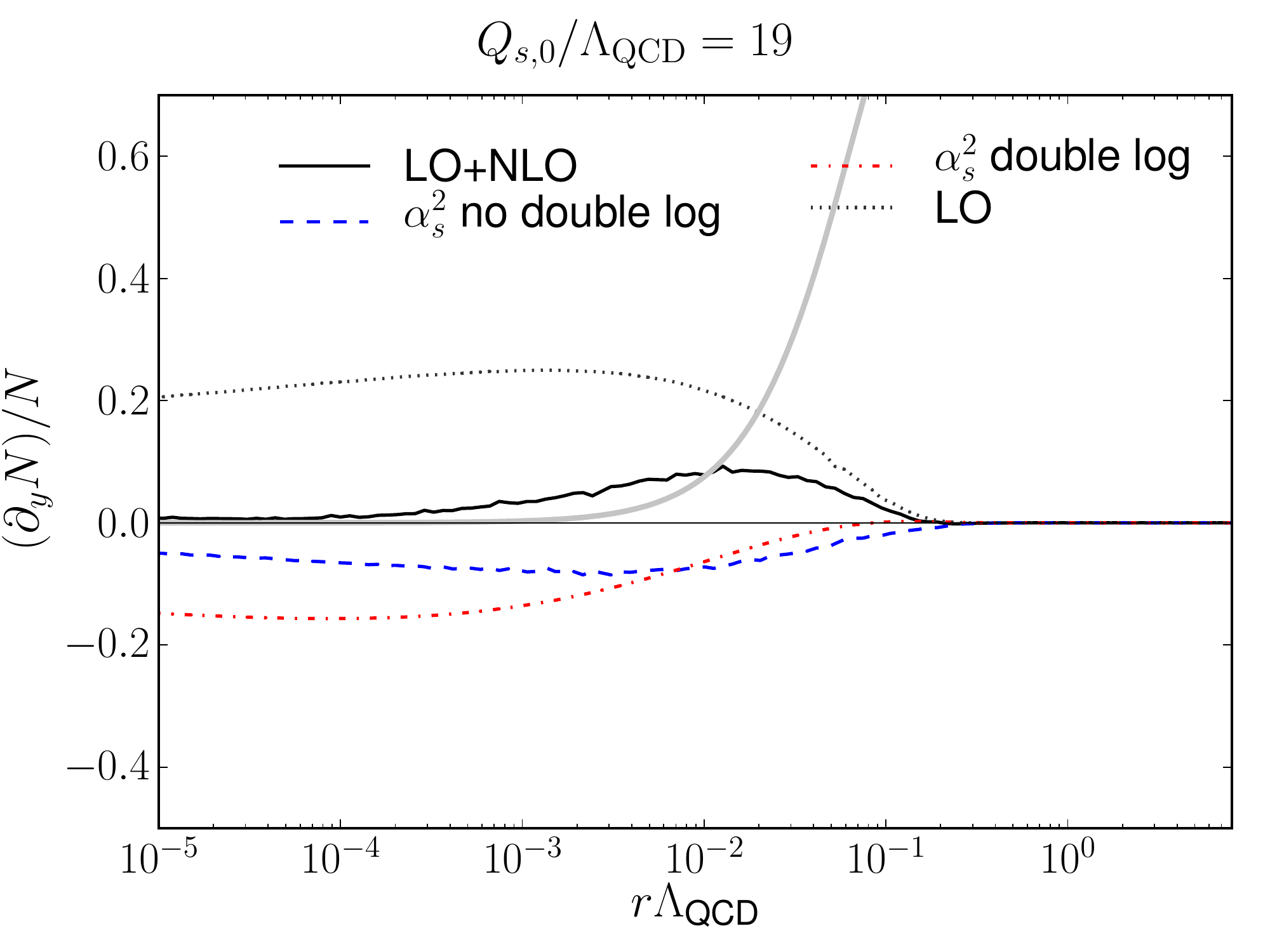}
		}
	\subfloat[$\gamma=1.0$]{
		\includegraphics[width=0.33\textwidth]{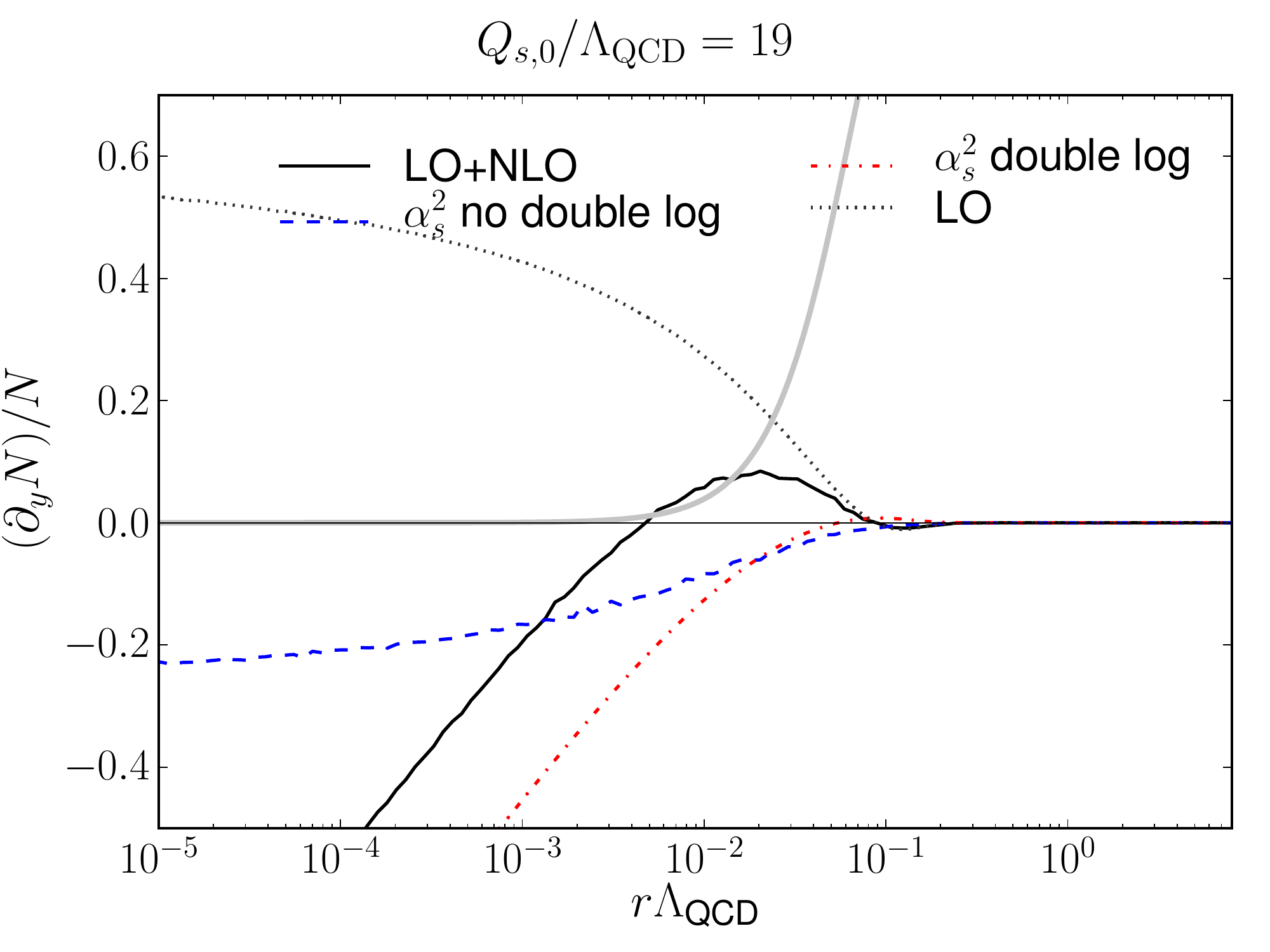}
		}			
	\caption{Evolution speed of the dipole amplitude at the initial condition. Shown are separately the full NLO and LO evolution equation results and the contributions from the conformal (no double logarithmic term) and non-conformal (only double logarithmic term) parts of the NLO BK equation. To demonstrate the location of the saturation scale the dipole amplitude is also shown as a thick line on the background.}
	\label{fig:dndy_qcd}
\end{figure*}

\begin{figure*}[ptb]
	\subfloat[$\gamma=0.6$]{
		\includegraphics[width=0.33\textwidth]{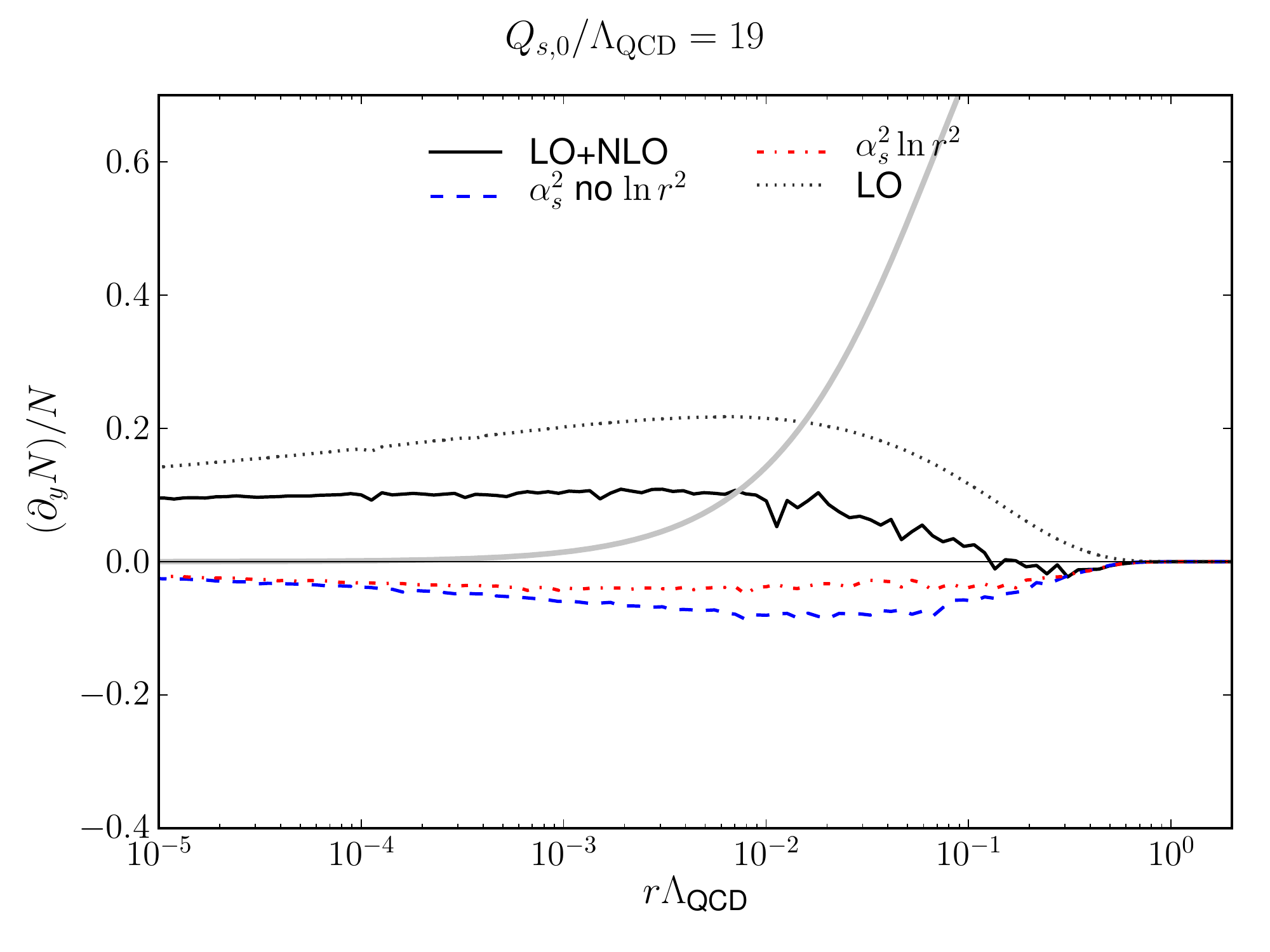}
		}
	\subfloat[$\gamma=0.8$]{
	\includegraphics[width=0.33\textwidth]{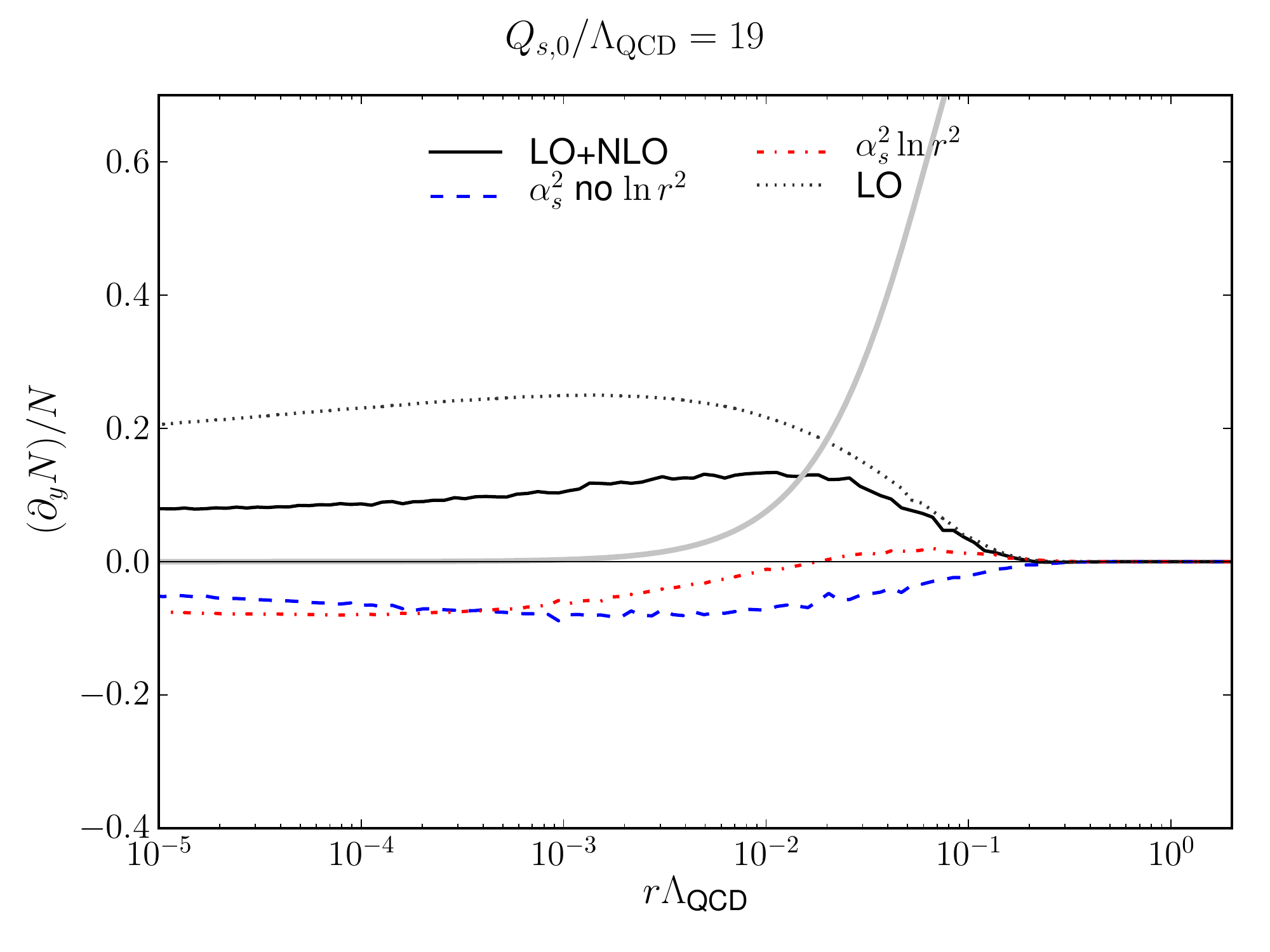}
		}
	\subfloat[$\gamma=1.0$]{
		\includegraphics[width=0.33\textwidth]{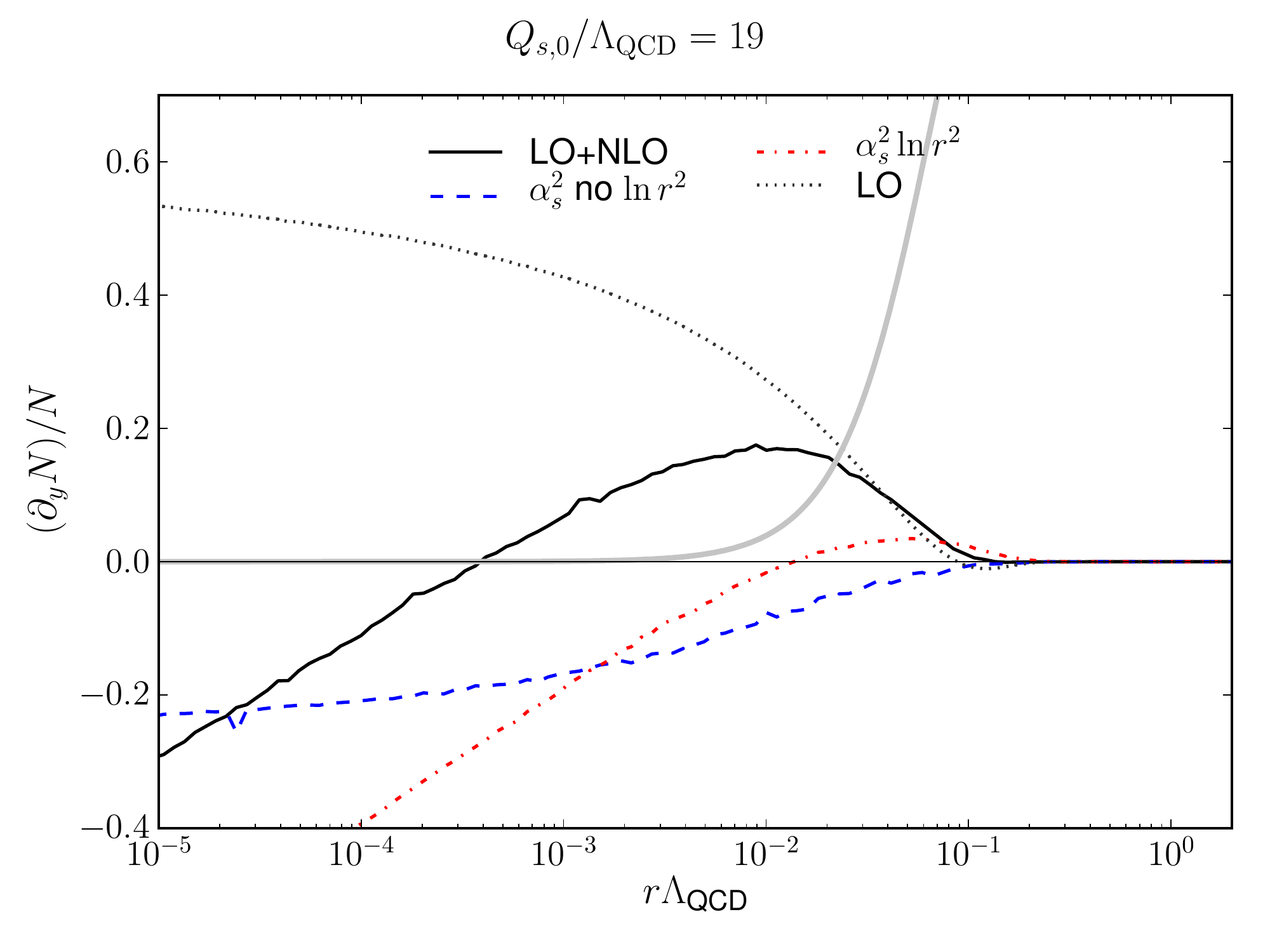}
		}	
	\caption{Evolution speed of the conformal dipole amplitude at the initial condition. Shown are separately the full NLO and LO evolution equation results and the contributions from the $\ln r^2$ term and the rest of the evolution equation. To demonstrate the location of the saturation scale the dipole amplitude is also shown as a thick line on the background.}
	\label{fig:dndy_confqcd}
\end{figure*}

We solve the evolution equations for the non-conformal and conformal dipoles on a logarithmical grid in $r$ using a Runge-Kutta method. The four-dimensional integral in the NLO part is computed using an adaptive Monte Carlo algorithm. As an initial condition we use a 
 McLerran-Venugopalan model~\cite{McLerran:1994ni}
\begin{equation}
	N(r) = 1 - \exp \left[ -\frac{(r^2 \qso^2)^\gamma}{4} \ln \left(\frac{1}{r \lqcd}+ e\right)\right],
\end{equation}
modified by introducing an anomalous dimension $\gamma$ which controls the power-like tail of the dipole amplitude for small dipoles. This parametrization is used in phenomenological fits to DIS data e.g. in Ref.~\cite{Albacete:2010sy}. 
Determining the correct values for $\qso^2$ and $\gamma$ would require a full NLO fit to e.g. DIS data, which we are not performing here. It is not obvious that the initial condition would be the same as for the leading order equation. We shall here merely explore the general behavior of the equation with different values for $\qso^2$ and $\gamma$ without aiming for phenomenologically relevant values in
this work.

We find that for some initial conditions (see discussion later) the evolution becomes unstable, 
such that the dipole amplitude starts to decrease and may even turn negative for small dipoles. It however follows from the definition of the dipole operator, \eq\eqref{eq:s-def}, that one should have $N(r)\to 0$ in the limit $r\to 0$, which is violated by non-zero amplitude at small $r$. Also the convolution with the kernel $K_1$ in \eq \eqref{eq:nlobk} does not converge if this requirement is not fulfilled. To obtain this property, we freeze $N(r)=0$ in the region where the evolution would turn it negative. 

The dipole amplitudes from the NLO equation are compared with the solution of the  leading order equation  in \fig\ref{fig:amplitude}. The main effect of the evolution is to increase the amplitude at small $r$, while maintaining it below the black disk limit of $N=1$ at large $r$. This leads to the curve $N(r)$ moving towards the left (smaller $r$) with rapidity in  \fig\ref{fig:amplitude}.
It can be seen that the NLO corrections reduce the evolution speed significantly but the shape of the dipole amplitude remains roughly unchanged. 
The solution in \fig\ref{fig:amplitude} has an initial condition
$\qso/\lqcd\sim 19,$ 
$\gamma=0.6$, deliberately
 chosen such that the dipole amplitude increases at small dipoles throughout the evolution over the rapidity interval studied here.
 The evolution for the conformal dipole with the same initial condition is also shown. Note that at leading order also the conformal dipole evolution is given by the standard LO BK equation. In the LO BK equation, we also use
the same ``Balitsky'' running coupling prescription. 
 The change of the saturation scale with energy is quantified more precisely  in \fig\ref{fig:dqsdy} with the evolution speed of the saturation scale 
\begin{equation}
 \lambda = \frac{\ud \ln \qs^2}{\ud y},
\end{equation}
where the precise definition of  $\qs^2$
used here is
\begin{equation}
\label{eq:qs}
	N(r^2=2/Q_s^2) = 1-e^{-1/2}.
\end{equation}
The NLO corrections can again be seen to significantly slow down the evolution speed. The conformal and ``non-conformal'' dipoles have comparable evolution speeds after a few evolution steps, and the total evolution speed decreases  slowly as a function of $\qs$. Note that the small anomalous dimension in the initial condition makes the leading order evolution faster than $\lambda\sim 0.2 \dots 0.3$ obtained in leading order fits with $\gamma \sim 1$~\cite{GolecBiernat:1998js,Albacete:2007sm,Albacete:2010sy,Lappi:2013zma}. Also the parameter $\qso^2$  that controls the initial saturation scale is not the same as the saturation scale $Q_s^2$ obtained by solving the equation \eqref{eq:qs}, and in this case $\qso/\lqcd\sim 19$ corresponds to having an initial saturation scale $Q_s/\lqcd \sim 40$.

One would generally expect $N$ to increase with rapidity, corresponding to the physical picture of more gluons being emitted when the available phase space increases with increasing collision energy.  This is the behavior seen in the LO equation.
To study when exactly this happens we show in \fig\ref{fig:dndy_dipole_y0} the evolution speed (logarithmic derivative of the dipole amplitude $\partial_y N(r)/N(r)$) at $y=0$ with different values for the anomalous dimension $\gamma$ and initial saturation scale $\qso$ as a function of the parent dipole size. We see that the scattering amplitude does indeed increase, but only for a suitable choice of the initial conditions: small enough $\gamma$ and large enough $\qso$.
Let us discuss the interpretation of the logarithmic derivative plots in more detail. 
For smaller $\qs$ the NLO corrections are so large that 
$\partial_y N(r)/N(r)$ is negative around the ``front'' $r\sim 1/\qs$, which makes the solution progress unphysically in the wrong direction, with $\qs$ decrasing with rapidity.
For larger $\qs$, the behavior around $r\sim 1/\qs$ is less problematic, and we can focus on the small $r$ tail of the amplitude. Here note that  if 
$\partial_y N(r)/N(r)$ has a constant positive value, the amplitude grows exponentially in rapidity, but retains its shape as a function of $r$, resembling the small $r$ behavior of the leading order evolution equation. 
This is indeed what happens for $\gamma=0.6$ and, marginally, for $\gamma=0.8$. For $\gamma=1.0$, however, we observe a negative, logarithmically decreasing
$\partial_y N(r)/N(r) \sim \ln r$ for $r \to 0$. This means that the evolution drives the 
amplitude towards a steeper shape, which in turn causes $\partial_y N(r)/N(r)$ to become even
more negative for small $r$. Eventually this unstable development leads to a singularity in 
$\partial_y N(r)/N(r)$, which means that $N(r)$ would cross zero at a finite $r$. 
At this point the integral on the r.h.s. of the BK equation would become divergent, so we impose $N(r)\geq0$ by hand. The same evolution speed at $y=5$ is shown in Fig.~\ref{fig:dndy_dipole_y5}, where it can be seen that the evolution speed remains sensitive to the details of the initial condition even at large rapidities.  The corresponding results for the conformal dipole are shown in Figs.~\ref{fig:dndy_confdipole_y0} and \ref{fig:dndy_confdipole_y5}, which show that the evolution speed of the conformal dipole is equally sensitive to the details of the initial condition.

To understand better this unstable behaviour of the dipole amplitude shape  we calculate  the anomalous dimension, defined as a logarithmic derivative of the dipole amplitude,
\begin{equation}
	\gamma(r) = \frac{\der \ln N(r)}{\der \ln r^2}.
\end{equation}
The anomalous dimensions for the conformal and non-conformal dipoles 
at different rapidities  are 
compared with the leading order BK solution in
\figs\ref{fig:gammaqcd} and \ref{fig:gammaconfqcd}. The instability of the solution is clearly visible at larger initial anomalous dimensions: with $\gamma=1$ in the initial condition the anomalous dimension at small dipoles grows very large already after a few rapidity steps. With $\gamma=0.8$, a significantly longer evolution is needed before the solution becomes unstable. When the initial anomalous dimension is smaller (here $\gamma=0.6$), the unstable region is not reached within the rapidity interval studied here. Note that the solution with $\gamma=1$ does not evolve significantly from $y=5$ to $y=30$, as the evolution is dominated by distance scales where the the dipole amplitude would have already become negative in the numerical solution and has been frozen to $N(r)=0$.

Contributions of different terms in the evolution equation~\nr{eq:nlobk}  are shown in Fig.~\ref{fig:dndy_qcd}  for the initial condition. In the figure the next-to-leading order contributions coming from the double logarithmic $\sim \as^2 \ln X^2/r^2 \ln Y^2/r^2$ and from the other $\sim \as^2$ terms are shown separately.  If the anomalous dimension in the initial condition is large enough, the double logarithmic term drives the evolution speed and is responsible for eventually turning the amplitude negative at small dipoles. With small $\gamma$ both NLO contributions approach zero in the small dipole limit. For large dipoles in the saturation regime the NLO contributions are still negative, but not large enough to change the sign of the evolution speed for $\qso/\lqcd = 19$. For a phenomenologically more relevant smaller value of $\qso$ the coupling is larger and the NLO corrections make the evolution speed negative for all dipole sizes, as seen in Figs.~\ref{fig:dndy_dipole_y0} and \ref{fig:dndy_confdipole_y0}.  

In the evolution equation of the conformal dipole, \eq \eqref{eq:conf-n}, the double logarithmic term is absent, but the definition of the conformal dipole makes an additional term $\sim \as^2 \ln r^2$ appear in the NLO evolution equation. 
It is now exactly this $\as^2 \ln r^2$ term which drives the conformal dipole negative at small dipoles if the anomalous dimension at the initial condition is large enough. This is clearly seen in \fig\ref{fig:dndy_confqcd}, where the NLO contributions coming from the $\sim \as^2  \ln r^2$ part and from the other NLO terms are shown. 
 Note that in the original NLO BK equation the only logarithm in addition to the problematic double logarithmic term is $\as^2 \ln X^2Y'^2/(X'^{2}Y^{2})$, which vanishes at $r=0$.
Compared to the evolution of the ``non-conformal'' dipole, the total evolution speed becomes negative at significantly smaller dipoles. If a smaller anomalous dimension is used at the initial condition, also the contribution from the $\ln r^2$ term vanishes for small dipoles.

In Ref.~\cite{Balitsky:2009xg} an evolution equation for the conformal dipole in $N=4$ SYM theory is derived. We have checked that the conformal dipole  in $N=4$ SYM has the same characteristic features as what was above shown in the case of QCD. Similarly we have also solved the evolution equation without using the Balitsky prescription for the running coupling in kernel $K_1$ but instead absorbing terms proportional to $\beta$ in the definition of $\as(\mu^2)$ and using the smallest dipole prescription for $\as$ by choosing the scale to be set by the smallest dipole, $\mu^2 = \min\{r^2,X^2,Y^2,X'^2,Y'^2,(z-z')^2 \}$. The characteristic features of the solutions do not change in this change of the running coupling prescription.

\section{Conclusions}

We have presented the first numerical solution to the next to leading order Balitsky-Kovchegov equation. The NLO corrections are shown to decrease the evolution speed and to be sensitive on the details of the initial condition.
The slower evolution speed obtained by solving the NLO evolution equations compared to the leading order BK is anticipated; in LO BK fits to HERA data the evolution speed needs to be  reduced by evaluating the running coupling at a higher scale; see discussion in Refs.~\cite{Lappi:2013zma,Lappi:2012vw}. However, as long as the solution to the evolution equation can become unphysical, too strong conclusions on the effect of the higher order corrections on the evolution speed can not be made.  

The fact that the dipole amplitude may, depending on the initial condition, become negative and non vanishing at small dipoles is unphysical. This problem is only partially cured by writing the equation in terms of the conformal dipole when the dipole amplitude becomes negative only at significantly smaller dipoles. Even though it is possible to obtain an evolution which satisfies the requirement of a vanishing dipole amplitude at zero dipole size limit, we would like to get a stable evolution also in the case $\qso\sim 1\gev, \gamma \sim 1$ which has been the relevant region for phenomenological fits using the leading order equation (note that the leading order fits prefer values $\gamma > 1$~\cite{Albacete:2010sy}).
We have shown that the problematic behavior of the equation is associated with large logarithms of the small parent dipole size $r$, which corresponds to large transverse momentum. This confirms the result of~\cite{Avsar:2011ds} and suggests that the BK equation would indeed require a resummation of the same contributions that have been discussed in the context of the NLO~BFKL equation~\cite{Salam:1998tj,Ciafaloni:1999yw,Altarelli:1999vw,Ciafaloni:2003rd}.
 This calls for a better understanding of the NLO evolution equation before the NLO dipole amplitude can be used in phenomenological applications.

\section*{Acknowledgements} 
We thank G. Chirilli, R. Paatelainen and G. Soyez for discussions. This work has been supported by the Academy of Finland, projects 
267321 and 273464, and the Graduate School of Particle and Nuclear Physics (H.M.)
and by computing resources from
CSC -- IT Center for Science in Espoo, Finland.

\bibliography{../../../refs}
\bibliographystyle{JHEP-2modM}

\end{document}